\documentclass[epj]{svjour}
\usepackage{amsfonts,amssymb,amsmath}
\usepackage{dcolumn,epsfig,color}

\newcommand{\fet}[1]{\mbox{\boldmath $#1$}}
\newcommand{\beq}{\begin{equation}}
\newcommand{\eeq}{\end{equation}}
\newcommand{\beqa}{\begin{eqnarray}}
\newcommand{\eeqa}{\end{eqnarray}}
\newcommand{\nn}{\nonumber \\ }
\begin{document}
\title{Towards high-order calculations of three-nucleon scattering in chiral effective field theory}
\author{E.~Epelbaum\inst{1} \and
  J.~Golak\inst{2} \and
  K.~Hebeler\inst{3}   \and
  H.~Kamada\inst{4}    \and
  H.~Krebs\inst{1} \and
  U.-G.~Mei{\ss}ner\inst{5,6,7}   \and
  A.~Nogga\inst{6}   \and
  P.~Reinert\inst{1} \and
  R.~Skibi\'nski\inst{2} \and
  K.~Topolnicki\inst{2} \and
  Y.~Volkotrub\inst{2} \and
  H.~Wita{\l}a\inst{2}
}                     
%
%
\institute{Ruhr-Universit\"at Bochum, Fakult\"at f\"ur Physik und
        Astronomie, Institut f\"ur Theoretische Physik II,  D-44780
        Bochum, Germany  \and
 M. Smoluchowski Institute of Physics, Jagiellonian
 University,  PL-30348 Krak\'ow, Poland \and
Institut f\"ur Kernphysik, Technische Universit\"at 
Darmstadt, 64289 Darmstadt, Germany \and
Department of Physics, Faculty of Engineering,
Kyushu Institute of Technology, Kitakyushu 804-8550, Japan \and
Helmholtz-Institut f\"{u}r Strahlen- und Kernphysik and Bethe Center for
Theoretical Physics, Universit\"{a}t Bonn, D-53115 Bonn, Germany \and
Institut f\"{u}r Kernphysik, Institute for Advanced~Simulation and
J\"{u}lich Center for Hadron Physics, \\ 
Forschungszentrum J\"{u}lich, D-52425 J\"{u}lich, Germany \and
Tbilisi State University, 0186 Tbilisi, Georgia
}
\date{Received: date / Revised version: date}
%
\abstract{We discuss the current status of chiral effective field
  theory in the three-nucleon sector and present selected results for 
  nucleon-deuteron 
  scattering observables based on
  semilocal momentum-space-regularized chiral two-nucleon potentials
  together with consistently regularized three-nucleon forces up to
  third chiral order. Using a Bayesian model for estimating
  truncation errors, the obtained results are found to provide a good
  description of the experimental data. We confirm our earlier
  findings that a high-precision description of nucleon-deuteron scattering
  data below pion production threshold will require the
  theory to be pushed to fifth chiral order. This conclusion
  is substantiated by an exploratory study of selected short-range
  contributions to the three-nucleon force at that order, which, as
  expected, are found to have significant effects on polarization
  observables at intermediate and high energies. We also outline the challenges
  that will need to be addressed in order to push the chiral expansion
  of three-nucleon scattering observables to higher orders. 
\PACS{
     {21.30.-x}{Nuclear forces}   \and
     {21.30.Fe}{Forces in hadronic systems and effective interactions}  \and
      {21.45.+v}{Few-body systems} 
     } 
} 
\maketitle
\section{Introduction}
\label{intro}

The past few years have seen remarkable advances towards pushing the
precision frontier of chiral effective field theory (EFT) in the two-nucleon
sector, see
Refs.~\cite{Epelbaum:2008ga,Epelbaum:2012vx,Machleidt:2011zz}
for review articles. The fifth-order (N$^4$LO) contributions to the nucleon-nucleon (NN)
force have been worked out in Ref.~\cite{Entem:2014msa} and a new generation of
accurate and precise chiral EFT NN potentials up through N$^4$LO has
been developed in
Refs.~\cite{Epelbaum:2014efa,Epelbaum:2014sza,Reinert:2017usi,Entem:2017gor},
see also Refs.~\cite{Ekstrom:2013kea,Gezerlis:2013ipa,Piarulli:2014bda} for related developments. 
The N$^4$LO interactions of Refs.~\cite{Reinert:2017usi,Entem:2017gor} utilize the
values of the pion-nucleon ($\pi$N) low-energy constants (LECs) determined
from matching chiral perturbation theory to the Roy-Steiner-equation
analysis of $\pi$N scattering at the subthreshold point \cite{Hoferichter:2015tha}, but
differ substantially by the regularization procedure. The potentials
developed by our group in Ref.~\cite{Reinert:2017usi} build upon our
earlier studies
\cite{Epelbaum:2014efa,Epelbaum:2014sza} and employ a local regulator
for pion-exchange contributions which, per construction, 
maintains the analytic structure of the long-range
interaction. Differently to the nonlocally regularized potentials of
Ref.~\cite{Entem:2017gor},
the interactions constructed in Ref.~\cite{Reinert:2017usi} do not
produce long-range
artifacts at any finite order of expansion in inverse powers of the momentum-space
cutoff $\Lambda$. The resulting semilocal\footnote{The term
  ``semilocal'' refers to a local regularization approach for
  long-range interactions in combination with a nonlocal cutoff
  for the short-range part of the nuclear force.}
momentum-space regularized (SMS) potentials of
Ref.~\cite{Reinert:2017usi} are currently the most precise chiral EFT
interactions and provide, at the highest order
N$^4$LO$^+$\footnote{The N$^4$LO$^+$ potentials of
  Ref.~\cite{Reinert:2017usi} include four sixth-order
  (i.e.~N$^5$LO) short-range operators contributing to F-waves, which
  are needed to describe certain very precisely measured proton-proton
  scattering observables at intermediate and high energies. The
  same operators are  included in the N$^4$LO potentials of
  Ref.~\cite{Entem:2017gor}.},  a nearly perfect and
$\Lambda$-independent (within the considered cutoff range)
description of neutron-proton and proton-proton scattering data below
pion production threshold from the self-consistent 2013 Granada
data base \cite{Perez:2013jpa}. The achieved description of the
scattering data is comparable to or even better than that
based on the so-called high-precision semi-phenomenological
potentials like the AV18 \cite{Wiringa:1994wb}, CD Bonn
\cite{Machleidt:2000ge} and Nijm I, II \cite{Stoks:1994wp} models.  

In spite of this exciting progress in the NN sector, applications of
chiral EFT to three- and more-nucleon systems and in processes
involving external sources are, with very few
exceptions, still limited to the next-to-next-to-leading order
(N$^2$LO) in the chiral expansion. What makes high-accuracy
calculations of such systems/reactions beyond N$^2$LO so difficult?  
The main technical and conceptual difficulties are related to the
treatment of many-body forces and exchange current operators.
Three- (3NF) and four-nucleon forces (4NF) start to contribute at third
(N$^2$LO) and fourth (N$^3$LO) orders of the chiral expansion,
respectively,  while
the first contributions to the exchange electroweak currents appear
already at second order (NLO) relative to the dominant single-nucleon
terms. Over the past decade, we have worked out off-shell-consistent
expressions for the 3NF \cite{Bernard:2007sp,Bernard:2011zr}, 4NF
\cite{Epelbaum:2006eu,Epelbaum:2007us} and 
electroweak charge and current operators
\cite{Kolling:2009iq,Kolling:2011mt,Krebs:2016rqz,Krebs:2019aka} completely up through N$^3$LO
using dimensional regularization (DR) to 
compute pion loop contributions, see also Refs.~\cite{Pastore:2008ui,Pastore:2009is,Piarulli:2012bn,Baroni:2015uza} for a related work on
electroweak currents by the JLab-Pisa group and Ref.~\cite{Hermann_review} for a
review article. Furthermore, selected N$^4$LO contributions to the 3NF
have been worked out  in
Refs.~\cite{Krebs:2012yv,Krebs:2013kha,Girlanda:2011fh}, and the
longest-range part of the 3NF was also analyzed in the framework
involving $\Delta(1232)$ degrees of freedom. On the technical side,
the implementation of the 3NFs and exchange currents in few-body
calculations requires their partial-wave decomposition. This
nontrivial task can nowadays be accomplished numerically for a general
3NF specified in momentum space as described in Refs.~\cite{Golak:2009ri,Hebeler:2015wxa}.
The main conceptual challenge that still needs to be addressed
concerns a {\it consistent} regularization of many-body forces and
exchange operators. While this issue is irrelevant in the NN sector, a
naive regularization of many-body interactions and exchange currents
based on the expressions derived using DR violates
chiral symmetry and leads to inconsistent/wrong results at the one-loop level
(i.e.~at N$^3$LO) and beyond \cite{CD18-Hermann,EE-NTSE}. A possible solution could be
provided by using higher-derivative regularization instead of DR,
which has to be chosen in the way compatible with the regularization
scheme of Ref.~\cite{Slavnov:1971aw} in the NN sector. Work along these lines is in
progress. 

In this paper we update our recent study \cite{Epelbaum:2018ogq} based on the
semilocal coordinate-space regularized (SCS) NN forces of Refs.~\cite{Epelbaum:2014efa,Epelbaum:2014sza} 
and analyze nucleon-deuteron (Nd) scattering observables using the
new SMS chiral NN potentials of Ref.~\cite{Reinert:2017usi} in combination with the SMS 3NF
at N$^2$LO. We also refine our previous estimations of truncation
uncertainties by employing a Bayesian approach instead of the
algorithm proposed in Ref.~\cite{Epelbaum:2014efa} and used in our earlier studies
\cite{Binder:2015mbz,Binder:2018pgl,Epelbaum:2018ogq} in the
three-nucleon sector. Last but nor least, we explore the role of
selected N$^4$LO short-range 3NF terms in Nd scattering.

Our paper is organized as follows. In section \ref{sec:2}, we specify our
Bayesian model for truncation errors by following the approach proposed
in  Ref.~\cite{Melendez:2017phj}. Our results for Nd scattering observables at
N$^2$LO are presented in section \ref{sec:3}, while the role of selected
N$^4$LO 3NF operators is discussed in section \ref{sec:4}. The main results
of our study are summarized in section \ref{sec:5}.

\section{Bayesian model for truncation errors}
\label{sec:2}

A reliable estimation of theoretical uncertainties is an essential ingredient
of any systematic approach such as chiral EFT. Cutoff
variation offers one possibility to quantify the impact of
contributions beyond the truncation level. However, in the few- and
many-nucleon sectors, the available cutoff range is often  rather
limited. Furthermore, cutoff variation does not allow one to probe the impact of
neglected long-range interactions.  
In Ref.~\cite{Epelbaum:2014efa}, a more reliable, universally applicable algorithm for estimating
truncation errors using the available information on the chiral
expansion for any observable of interest  
without relying on cutoff variation was proposed. Here and in what
follows, this algorithm will be referred to as the EKM-approach. For
applications of the EKM method to a broad range of low-energy reactions
in the single-baryon  and few-/many-nucleon sectors see
Refs.~\cite{Siemens:2016hdi,Yao:2016vbz,Siemens:2017opr,Blin:2018pmj}
and
\cite{Epelbaum:2014sza,Binder:2015mbz,Hu:2016nkw,Skibinski:2016dve,Binder:2018pgl,Epelbaum:2018ogq,NevoDinur:2018hdo},
respectively. Being very simple and easy to implement, the EKM
approach does, however, not directly provide a statistical
interpretation of the estimated uncertainties. In
Refs.~\cite{Furnstahl:2015rha,Melendez:2017phj}, a general Bayesian approach
to calculate the posterior probability distribution for truncation
errors in chiral EFT was developed. The EKM
approach was then shown to essentially correspond to a particular
choice of prior probability distribution for the coefficients in the
chiral expansion of observables. Using the chiral NN potentials from
Ref.~\cite{Epelbaum:2014efa}, the EKM error estimations for the
neutron-proton (np) total cross
section at selected energies were found in the Bayesian approach of Ref.~\cite{Furnstahl:2015rha} to be
consistent with $68\%$  degree-of-belief (DoB) intervals. 

Throughout this paper, we employ a slightly modified version of the
Bayesian model from Ref.~\cite{Furnstahl:2015rha}. Specifically,
consider a two-nucleon scattering observable $X(p)$ with $p$ referring
to the center-of-mass (CM) momentum. Calculating $X(p)$ using chiral EFT
potentials at various orders $Q^i$, $i=0, 2, 3, 4,
\ldots$,  (but for a fixed cutoff value) yields the corresponding
predictions $X^{(i)} (p)$, and the chiral expansion of $X(p)$ can be
written in the form
\beqa
\label{coeffci}
X &=& X^{(0)} + \Delta X^{(2)} + \Delta X^{(3)} + \Delta X^{(4)} +
\ldots \nn
&=:& X_{\rm ref} \left( c_0 + c_2 Q^2 + c_3 Q^3 + c_4 Q^4 + \ldots \right) \,,
\eeqa
with $\Delta X^{(2)} := X^{(2)} - X^{(0)}$ and   $\Delta X^{(i)} :=
X^{(i)} - X^{(i-1)}$ for $i > 2$.  The second equality serves as a
definition of the dimensionless expansion coefficients $c_i$. The
reference value $X_{\rm ref}$ which sets the overall scale will be
defined below. Here and in what follows, the
expansion parameter of chiral EFT is assumed to have the form
\cite{Epelbaum:2014efa,Epelbaum:2014sza,Furnstahl:2015rha,Melendez:2017phj}  
\begin{equation}
  \label{ExpParam}
Q = \max \left( \frac{p}{\Lambda_b}, \, \frac{M_\pi^{\rm
      eff}}{\Lambda_b} \right)\,,
\end{equation}
 where $\Lambda_b$ is the breakdown scale of the chiral expansion.
The
quantity $M_\pi^{\rm  eff}$ serves to model the expansion of NN
observables around the chiral limit. In
Refs.~\cite{Epelbaum:2014efa,Epelbaum:2014sza,Furnstahl:2015rha,Melendez:2017phj},
this scale was
set to the physical pion mass $M_\pi$. However, as pointed out in Ref.~\cite{CD18-EE}, the
error plots in Ref.~\cite{Epelbaum:2014efa} indicate that the transition between the
two expansion regimes in the NN sector actually appears at a scale $M_\pi^{\rm
  eff}$  higher than $M_\pi$.  On the other hand, both Bayesian model parameters $\Lambda_b$ and
$M_\pi^{\rm  eff}$ can be determined/tuned empirically by
calculating the success rates for a given set of observables and/or
energies. In particular, Ref.~\cite{Furnstahl:2015rha} confirmed the EKM
estimation 
$\Lambda_b \sim 600$~MeV based on the results for the total np cross
section and using the SCS potentials of Ref.~\cite{Epelbaum:2014efa},
but it also found somewhat larger values
of $\Lambda_b$ to be statistically consistent, see also a related
discussion in \cite{Melendez:2019izc}.  A similar empirical analysis
was performed in Ref.~\cite{CD18-EE} for both $\Lambda_b$ and $M_\pi^{\rm
  eff}$ yielding the values of $\Lambda_b \sim 650 \ldots 700$~MeV
and $M_\pi^{\rm eff} \sim 200$~MeV.  

Suppose the results for the observable $X(p)$ are available up through
the order $X^{(k)}$, $k \ge 2$. The goal is then to estimate the
truncation error $\delta X^{(k)} \equiv \sum_{i > k} \Delta X^{(i)}$
resulting from neglecting the unknown higher-order contributions,
i.e.~to compute the posterior probability distribution function (pdf)
for $\delta X^{(k)}$ given the explicit knowledge of $\{ X^{(0)}, \, X^{(2)},
\, \ldots , X^{(k)} \}$. The Bayesian model of
Ref.~\cite{Melendez:2017phj} uses the leading-order (LO) result for $X(p)$ to
set the overall scale
\beq
\label{LOscale}
X_{\rm ref} = X^{(0)}
\eeq
in order to define the
dimensionless expansion coefficients $c_i$ with $c_0=1$  in
Eq.~(\ref{coeffci}).
As we will argue below, this approach may, for certain choices of the
prior pdf, be too restrictive in
the kinematical regions near the points where the LO contribution
vanishes e.g.~by changing the sign. Such situations
are not uncommon if one looks at observables which depend on
continuously varying parameters such as energy or scattering angle,
see also a discussion in Ref.~\cite{Melendez:2019izc}. In such
circumstances it is advantageous to set the overall scale from the 
next-to-leading order (NLO) contribution $\Delta X^{(2)}$ via $X_{\rm
  ref} = \Delta X^{(2)}/Q^2$ in order to avoid underestimating $X_{\rm
  ref}$. To have an approach applicable when both $X^{(0)}$ and/or
$\Delta X^{(2)}$ are accidentally small, we set the scale $X_{\rm
  ref}$ via
\beq
\label{LONLOscale}
X_{\rm ref} = \max \left( | X^{(0)} |, \;  \frac{| \Delta X^{(2)}
    |}{Q^2} \right)
\eeq
for $k = 2$ and
\beq
\label{LONLON2LOscale}
X_{\rm ref} = \max \left( | X^{(0)} |, \;  \frac{| \Delta X^{(2)}
    |}{Q^2} , \;  \frac{| \Delta X^{(3)}
    |}{Q^3} \right)
\eeq
for $k \ge 3$. Let $c_m = 1$, $m \in \{0, 2, 3\}$, be the coefficient used
to define the overall scale. Assuming that the remaining
coefficients $c_i$ are distributed according to some common pdf ${\rm
  pr} (c_i | \bar c )$ with a hyperparameter $\bar c$
and performing marginalization over $h$
chiral orders $k+1, \ldots, k+h$, which are assumed to dominate the truncation
error,
the probability distribution for the
dimensionless residual $\Delta_k \equiv \sum_{n=k+1}^\infty c_n Q^n
\simeq  \sum_{n=k+1}^{k+h} c_n Q^n$
to take a value $\Delta_k = \Delta$, given the knowledge of $\{c_{i
  \le k} \}$, is given by \cite{Melendez:2017phj}
\begin{equation}
  \label{posteriorGeneral}
{\rm pr}_h ( \Delta  | \{ c_{i \le k} \}) = \frac{\int_0^\infty d \bar c \,
  {\rm pr}_h (\Delta | \bar c ) \,  {\rm pr} (\bar c ) \prod_{i\in A}
  {\rm pr} (c_i | \bar c ) }{\int_0^\infty  d \bar c \,
  \prod_{i\in A}  {\rm pr} (c_i | \bar c ) }\,,
\end{equation}
where the set $A$ is defined as $A = \{n
\in \mathbb{N}_0 \, | \, n \leq k \, \land \, n \neq 1 \, \land \, n \neq m \}$ and
 \begin{equation}
 {\rm pr}_h (\Delta | \bar c ) \equiv \left[ \prod_{i=k+1}^{k+h}
   \int_{-\infty}^\infty d c_i  {\rm pr} (c_i | \bar c ) \right]
 \, \delta \bigg( \Delta - \sum_{j=k+1}^{k+h} c_j Q^j \bigg)\,.
 \end{equation}
Here and in what follows, we employ the 
Gaussian prior of ``set C'' from Ref.~\cite{Melendez:2017phj}, namely
\begin{equation}
  \label{prior}
 {\rm pr} (c_i | \bar c ) = \frac{1}{\sqrt{2 \pi} \bar c} \, e^{-
   c_i^2/(2 \bar c^2 )}
\end{equation}  
and assume a log-uniform probability distribution \cite{Schindler:2008fh}
\begin{equation}
{\rm pr} ( \bar c ) = \frac{1}{\ln ( \bar c_> / \bar c_< )} \, 
\frac{1}{\bar c} \, \theta (\bar c - \bar c_< ) \, \theta (\bar c_> -
\bar c )\,.
\end{equation}
The nice feature of this prior is that all integrations in
Eq.~(\ref{posteriorGeneral}) can be carried out analytically. For the
sake of completeness, we give in Appendix \ref{sec:app1} the
corresponding expressions for ${\rm pr}_h ( \Delta  | \{ c_{i \le k} \})$ from Ref.~\cite{Melendez:2017phj}. 
The posterior pdf is an even function of $\Delta$, and for any given
DoB interval, the corresponding value of the residual $\Delta_k$ and
the resulting truncation error $\delta X^{(k)} = X_{\rm ref} \Delta_k$
can be readily obtained by
numerically integrating ${\rm pr}_h ( \Delta  | \{ c_{i \le k} \})$ over
$\Delta$.  

\begin{figure}[tb]
\includegraphics[width=0.49\textwidth,keepaspectratio,angle=0,clip]{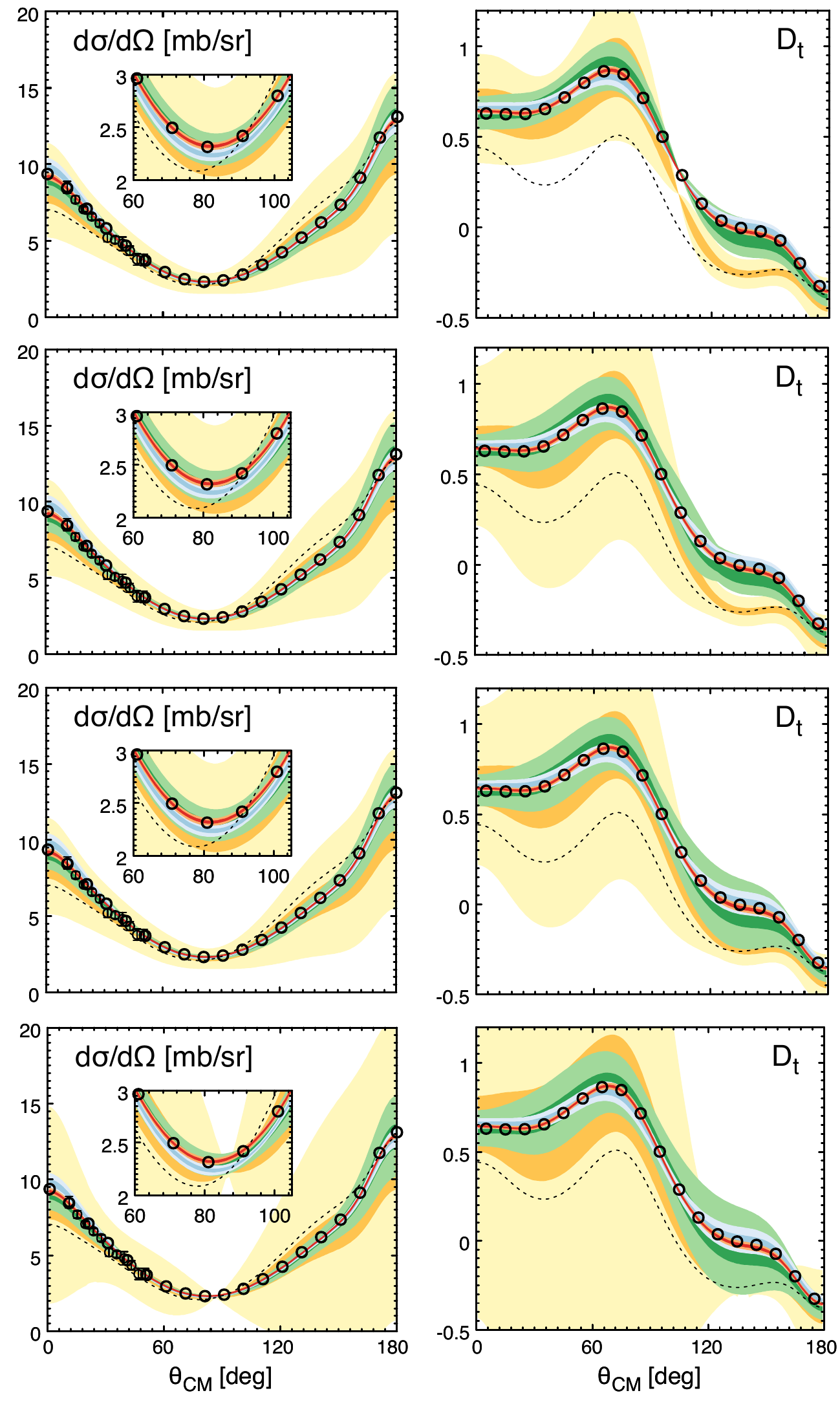}
\caption{Estimated theoretical uncertainty for the chiral EFT results
  for np differential cross section $d \sigma/d \Omega$ (left panel)
  and polarization transfer coefficient $D_t$ at laboratory energy of
  $E_{\rm lab} = 143$~MeV. The light- (dark-) shaded
yellow, green, blue and red bands of decreasing width depict $95\%$ ($68\%$)
DoB intervals at NLO, N$^2$LO, N$^3$LO and N$^4$LO, respectively.
Dashed lines show the LO predictions.
Open
circles refer to the results of the Nijmegen partial wave analysis
\cite{Stoks:1993tb}. Data for the cross section are at $E_{\rm lab} = 142.8\,$MeV and taken from \cite{ber76}.
The first, second, third and fourth rows correspond to the Bayesian
models $C_{0.5-10}^{650}$,  $\tilde C_{0.5-10}^{650}$, $\bar
C_{0.5-10}^{650}$ and $C_{\epsilon}^{700}$.  All results shown are
based on the SMS NN potentials using the cutoff of $\Lambda = 450$~MeV. 
}
\label{fig:02}       
\end{figure}

\begin{figure}[tb]
\includegraphics[width=0.49\textwidth,keepaspectratio,angle=0,clip]{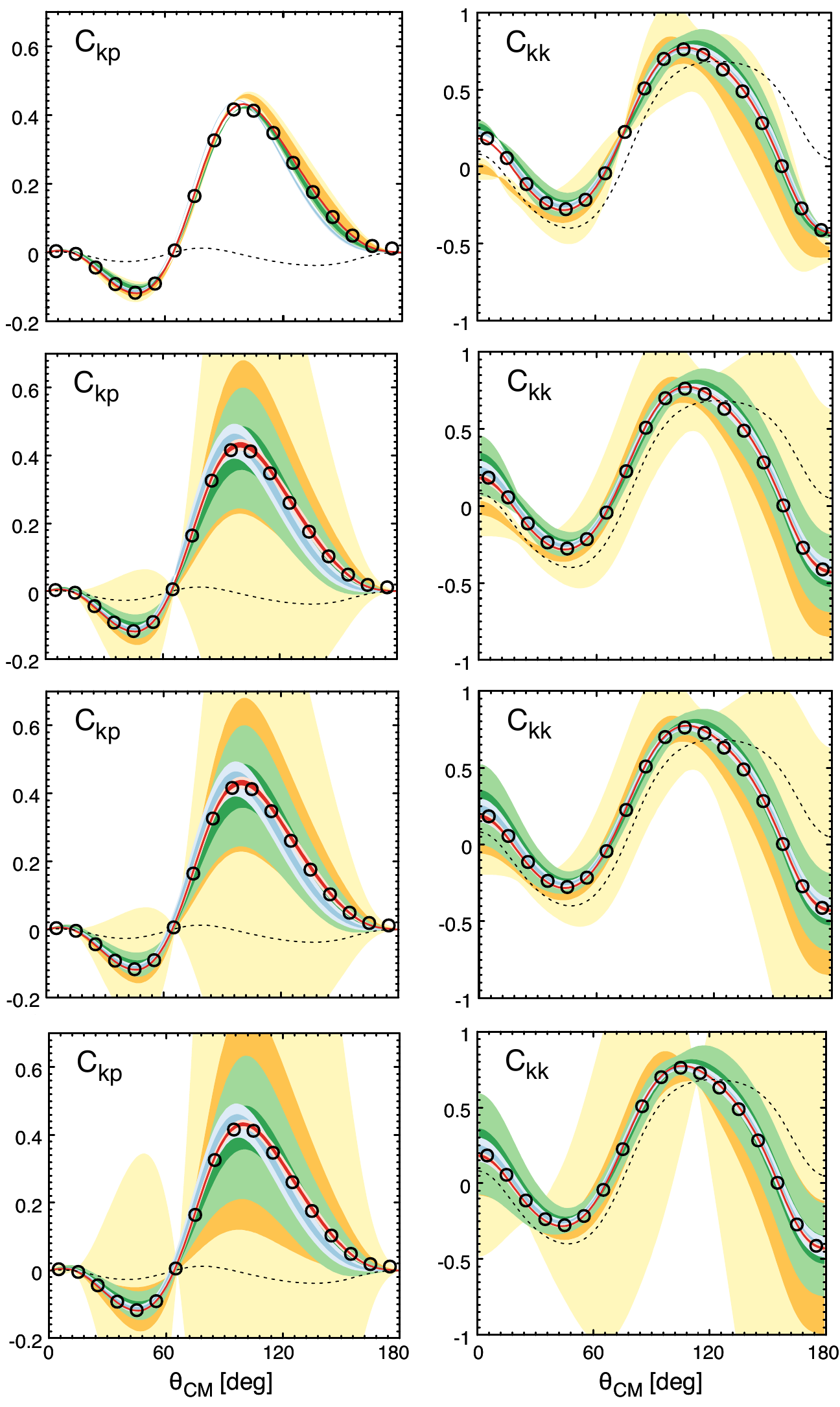}
\caption{Same as Fig.~\ref{fig:02} but for the spin-correlation
  parameters $C_{kp}$ and $C_{kk}$.}
\label{fig:03}       
\end{figure}

As a first application, we employ the Bayesian model of
Ref.~\cite{Melendez:2017phj} and set the overall scale $X_{\rm ref}$
solely from the corresponding LO contribution, see Eq.~(\ref{LOscale}).  
In the bottom row of Figs.~\ref{fig:02} and \ref{fig:03}, we show the
$68\%$ and $95\%$ DoB intervals for selected np scattering observables
at $E_{\rm lab}= 143$~MeV for the non-informative prior
$C_\epsilon$ \cite{Furnstahl:2015rha} with $\bar c_< = \epsilon$
and $\bar c_> = 1/\epsilon$, which makes no assumption about either the
maximum and minimum size of $\bar c$ by taking the limit $\epsilon \to
0$, see Eq.~(\ref{priorepsilon}). Here, the breakdown scale was set
to $\Lambda_b = 700$~MeV, and the resulting Bayesian model is referred
to as $C_\epsilon^{700}$.     
Being noninformative, the prior set $C_\epsilon$ is generally expected
to yield conservative estimates for truncation errors. However,
setting $\bar c_< \to 0$ yields a $\delta$-function-like posterior ${\rm
  pr}_h^{C_\epsilon} ( \Delta | \{ c_{i \le k}\} )$ for $\fet c_k^2
\to 0$ as can be seen from Eq.~(\ref{priorepsilon}), i.e.~this model
fails to provide an adequate estimation of the truncation error if the
corrections $\Delta^{(2)}, \; \ldots , \; \Delta^{(k)}$ 
happen to be accidentally small. For the examples shown in
Figs.~\ref{fig:02} and \ref{fig:03}, this is the case at NLO for the
differential cross section $d \sigma/d \Omega$ around $\theta_{\rm CM}
\sim 85^\circ$, for the polarization transfer $D_t$ around $\theta_{\rm CM}
\sim 140^\circ$, for the spin correlation coefficients $C_{kp}$ at $\theta_{\rm CM}
\sim 65^\circ$ and for the coefficient $C_{kk}$ at $\theta_{\rm CM}
\sim 25^\circ$, and $\theta_{\rm CM}
\sim 115^\circ$. In all these cases, the corresponding functions
$\Delta X^{(2)} (\theta_{\rm CM} )$ change their sign. To circumvent
the problem with the underestimation of the truncation error in such
kinematical regions, the authors of Ref.~\cite{Melendez:2017phj} suggested to
use a more informative (but not too restrictive) prior set
$C_{0.25-10}$ corresponding to $\bar c_< = 0.25$ and $\bar c_> = 10$. Here and
in what follows, we make the choice $\bar c_< = 0.5$, which we found  to be
more efficient in resolving the above mentioned issue while still
sufficiently general.  As shown in the upper row of Figs.~\ref{fig:02}
and \ref{fig:03}, the prior set $C_{0.5-10}^{650}$ indeed yields
reasonable estimates of the  truncation errors for $d \sigma/d
\Omega$.
However, the more
informative prior with $\bar c_< \neq 0$ suffers from another issue as it
yields a vanishingly small truncation error 
at all orders in the cases when $X^{(0)}$ happens to be 
accidentally small. This is the case for $D_t$ at $\theta_{\rm CM}
\sim 100^\circ$ and for $C_{kk}$ at $\theta_{\rm CM}
\sim 10^\circ$, $\theta_{\rm CM}
\sim 75^\circ$ and $\theta_{\rm CM}
\sim 180^\circ$,  see the plots in the upper row of Figs.~\ref{fig:02}
and \ref{fig:03}. The most extreme situation is observed for the
coefficient $C_{kp}$, for which the LO contribution appears to be
small for all scattering angles.  The problem can be
traced back to the misidentification of the overall scale by
Eq.~(\ref{LOscale}) in such accidental cases. Writing $X^{(0)}$ as  $ X^{(0)}=
\alpha \tilde X^{(0)} $ with $\alpha \to 0$ being a dimensionless
parameter, one finds $\Delta_k \sim \alpha^{-1} $ for the prior set $C_\epsilon$ while 
 $\Delta_k \sim \alpha^0$  leading to 
$\delta X^{(k)} = \alpha \tilde X^{(0)} \Delta_k \sim \alpha $
for the prior set $C_{\bar c_< -  \bar c_>}$ with $\bar c_> < \infty$.  
The problem can be easily fixed by replacing Eq.~(\ref{LOscale}) with
Eq.~(\ref{LONLOscale}) as shown in the second row of Figs.~\ref{fig:02}
and \ref{fig:03}.  Here and in what follows, the resulting
Bayesian model is referred to as $\tilde C$.
However, while highly unlikely, it is still possible that both the LO contribution and the NLO
correction are simultaneously accidentally small. For the considered observables, this
happens for  $D_t$ at backward and for $C_{kk}$ at forward angles. To
prevent underestimating the truncation errors
in such rare cases, we replace, for $k \geq 3$, Eq.~(\ref{LONLOscale}) with 
Eq.~(\ref{LONLON2LOscale}) and refer to the resulting Bayesian model
as $\bar C$. This model is found to yield
robust results for NN scattering observables in all kinematical regions, see e.g.~the third
row in Figs.~\ref{fig:02} and \ref{fig:03} and will be employed in
nucleon-deuteron scattering calculations considered in the next
sections. 

\begin{figure}[tb]
\includegraphics[width=0.49\textwidth,keepaspectratio,angle=0,clip]{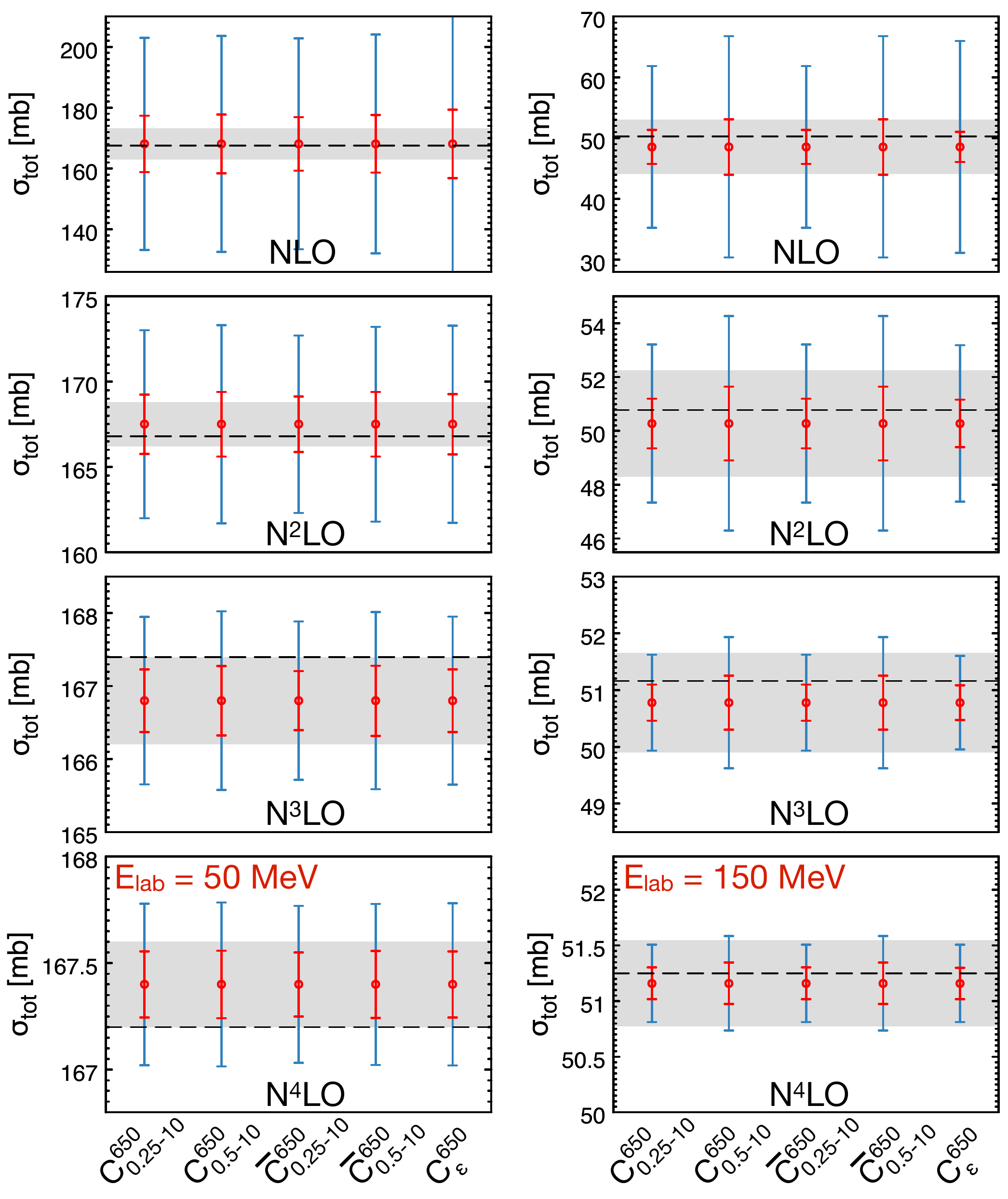}
\caption{(Color online) Neutron-proton cross sections at $E_{\rm lab}
  = 50$~MeV (left panel) and $E_{\rm lab}
  = 150$~MeV (right panel) at various orders of the chiral expansion
  using the SMS potentials of Ref.~\cite{Reinert:2017usi} with the
  cutoff $\Lambda = 450$~MeV. The smaller
  (red) error bars correspond to $68\%$ DoB intervals while the larger
  (blue) ones indicate  $95\%$ DoB intervals for a variety of Bayesian
  models as described in the text. The gray shaded bands show the
  uncertainty estimates using the original EKM approach. Dashed lines
  show the result of the next-order calculation (N$^2$LO at NLO,
  N$^3$LO at N$^2$LO, N$^4$LO at N$^3$LO and N$^4$LO$^+$ at N$^4$LO).
} 
\label{fig:01}       
\end{figure}

It is important to emphasize that the considered examples have been
selected to visualize the possible issues in certain accidental
situations, and that the differences between the considered Bayesian
models are fairly minor for most observables. This is exemplified in Fig.~\ref{fig:01},
where the truncation errors for the total cross section are shown for
a variety of considered models including the original EKM-approach
(with $M_\pi^{\rm eff} = M_\pi$ and $\Lambda_b = 600$~MeV).  
As pointed out in Ref.~\cite{Furnstahl:2015rha}, the dependence on a particular
Bayesian model
and/or assumed prior set 
decreases with an increasing order, i.e.~with the increasing amount of
information about the actual pattern of the chiral expansion. Notice
that differently to the EKM approach, the considered Bayesian models
exploit only the information up to the order, at which the truncation
error is estimated.  The more
conservative error estimations at $E_{\rm lab} = 50$~MeV with the
models $C$ and $\bar C$ as compared to the original EKM approach are mainly due
to the larger value of $M_\pi^{\rm eff}$. For the considered
set of the total cross section calculations, counting the success rate for the next-higher order result
to lie within the estimated uncertainty as shown in Fig.~\ref{fig:01} yields
the values of $62.5\ldots 75\%$ ($100\%$), which are
statistically consistent with the DoB intervals of $68\%$ 
($95\%$).\footnote{For the EKM approach, the success rate
  equals $100\%$ 
  per construction.}   We have verified that this
conclusion also holds true for a larger set of energies considered in
Ref.~\cite{CD18-EE}. We refrain from performing similar statistical tests for
angular distributions due to their correlated nature. This could be
done using the method proposed in Ref.~\cite{Melendez:2019izc}, which is based on
Gaussian processes and encodes
correlation structure of coefficients $c_i (\theta_{\rm CM} )$.    

\section{Nucleon-deuteron scattering at N$^2$LO using SMS nuclear potentials}
\label{sec:3}

We now turn to the main topic of our study and consider
nucleon-deuteron scattering in chiral EFT. 
The N$^2$LO three-nucleon force is given by \cite{vanKolck:1994yi,Epelbaum:2002vt}
\beqa
\label{leading}
V^{\rm 3N} &=& \frac{g_A^2}{8 F_\pi^4}\; 
\frac{\vec \sigma_1 \cdot \vec q_1  \; \vec \sigma_3 \cdot \vec q_3 }{[
 \vec  q_1^2 + M_\pi^2] \, [\vec q_3^2 + M_\pi^2]} \;\Big[ \fet \tau_1
\cdot \fet \tau_3  \big( - 4 c_1 M_\pi^2 \nn
&+& 2 c_3 \, \vec q_1 \cdot \vec
    q_3 \big)  
+  c_4 \fet \tau_1
  \times \fet \tau_3  \cdot \fet \tau_2  \; \vec q_1 \times \vec q_3 
\cdot \vec \sigma_2  \Big]  \nn
&-& \frac{g_A \, D}{8 F_\pi^2}\;  
\frac{\vec \sigma_3 \cdot \vec q_3 }{\vec q_3^2 + M_\pi^2} \; 
\fet \tau_1 \cdot \fet \tau_3 \; \vec \sigma_1 \cdot \vec q_3 \; + \; 
\frac{1}{2} E\, \fet \tau_1 \cdot \fet \tau_2\nn
&+&  \mbox{5 permutations}\,, 
\eeqa
where the subscripts refer to the nucleon labels, $\vec q_{i} = \vec p_i \,
' - \vec p_i$  with $\vec p_i \, '$
and $\vec p_i$ being the final and initial momenta of the nucleon $i$ 
and  $\vec \sigma_i$  ($\fet \tau_i$) are the Pauli
spin (isospin) matrices. Further,  
$c_i$, $D$ and $E$ denote the corresponding low-energy constants
(LECs) while $g_A$ and $F_\pi$ refer to the
nucleon axial coupling and pion decay constant, respectively. It is customary to
express the LECs $D$ and $E$
in terms of the corresponding dimensionless
coefficients via
\beq
\label{DimLessDE}
D = \frac{c_D}{F_\pi^2 \Lambda_\chi}, \quad \quad
E = \frac{c_E}{F_\pi^4 \Lambda_\chi} \,,
\eeq
where, following \cite{Epelbaum:2002vt,Epelbaum:2018ogq}, we use $\Lambda_\chi = 700$~MeV.  
For the LECs $c_i$, we
employ the central values from the Roy-Steiner-equation analysis of
Ref.~\cite{Hoferichter:2015tha} at the corresponding chiral order, namely $c_1=
-0.74$~GeV$^{-1}$, $c_3 = -3.61$~GeV$^{-1}$ and $c_4 = 2.44$~GeV$^{-1}$. The same values
are used in the SMS NN potentials of Ref.~\cite{Reinert:2017usi} at
N$^2$LO. Differently to Ref.~\cite{Epelbaum:2018ogq}, we employ the
same semilocal momentum-space  regulator for the 3NF as in
the  NN potentials of Ref.~\cite{Reinert:2017usi} by replacing the
pion propagators via 
\beq
\label{regLoc}
\frac{1}{\vec q_i^2 + M_\pi^2 } \; \; \to \; \; \frac{1}{\vec q_i^2 +
  M_\pi^2 } e^{- \frac{\vec q_i^2 +
M_\pi^2}{\Lambda^2}}\,.
\eeq
For the $D$-term, the contact interaction between the
nucleons $1$ and $2$ is, in addition, regularized by multiplying the 
matrix elements with a nonlocal Gaussian regulator $\exp(- (\vec p_{12}^2 +
  {\vec p_{12}'} ^2)/\Lambda^2 )$, where $\vec p_{12} = (\vec p_1 - \vec
  p_2 )/2$, ${\vec p}_{12}^{\, \prime} = (\vec p_1^{\, \prime} - \vec
  p_2^{\, \prime} )/2$.  For the contact interaction proportional to the LEC $E$, we apply a nonlocal
 Gaussian  regulator in momentum space 
  \beq
  \label{RegulSR}
V^{3N}_{\rm cont}  \; \;  \to  \; \; V^{3N}_{\rm cont}  \,
\; e^{ -\frac{4 \vec p_{12}^2 + 3 \vec k_3^2}{4 \Lambda^2}
  } \; 
e^{ -\frac{4 {\vec p_{12}'}^{2} + 3 {\vec k_3'}^2}{4 \Lambda^2} } \,,
\eeq
where $\vec k_3 = 2 (\vec p_3 - (\vec p_1 + \vec p_2)/2)/3$ and $\vec
k_3^{\, \prime} = 2 (\vec p_3^{\, \prime} - (\vec p_1^{\, \prime} + \vec p_2^{\, \prime})/2)/3$ are
the corresponding Jacobi momenta. 

It is important to emphasize that the SMS NN potentials of Ref.~\cite{Reinert:2017usi}
employ additional (local) short-range subtractions to ensure that the
coordinate-space expressions of the regularized pion-exchange
contributions and derivatives thereof vanish at the origin. This
convention  ensures that regularized pion-exchange contributions
contain only long-range pieces. On the other hand, using
the regulator in Eq.~(\ref{regLoc}), the resulting TPE contributions
still contain admixtures of short-range terms of the $D$-
and $E$-types.

Following Ref.~\cite{Epelbaum:2018ogq}, we determine the LECs $c_D$ and
$c_E$ from the $^3$H binding energy and the nucleon-deuteron
differential cross section minimum at $E_{\rm lab}^N =
70$~MeV. Specifically, we fit to the experimental data in the range of
$\theta_{\rm CM} = 107\ldots 141^\circ$.    The resulting $c_D$- and
$c_E$-values are strongly dependent on the cutoff $\Lambda$ and change
from $c_D = 8.9$ ($c_E = 1.15$)  for $\Lambda = 400$~MeV to $c_D = -5.4$
($c_E = -0.25$) for $\Lambda = 550$~MeV. Such a strong cutoff
dependence has to be expected given
that these LECs refer to bare quantities. The calculated observables,
on the other hand, show only a weak residual cutoff dependence
consistent with the estimated truncation uncertainty. 

\begin{figure}[tb]
\includegraphics[width=0.49\textwidth,keepaspectratio,angle=0,clip]{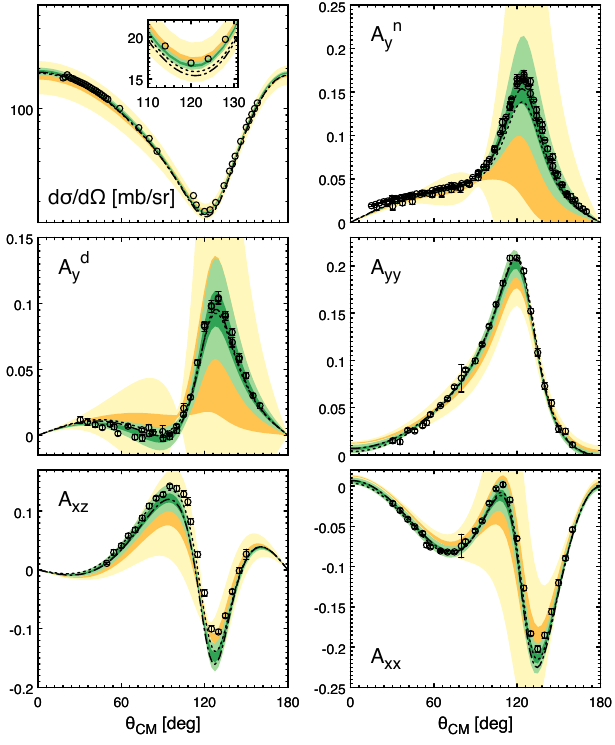}
\caption{Predictions for the differential cross section, nucleon and
  deuteron analyzing powers $A_y^n$ and $A_y^d$ as well as deuteron
  tensor analyzing powers $A_{yy}$, $A_{xz}$ and $A_{xx}$ in elastic
  nucleon-deuteron scattering at laboratory energy of
  $E_{\rm lab}^N = 10$~MeV at NLO
  (yellow bands) and N$^2$LO (green bands) based on the SMS NN
  potentials of Ref.~\cite{Reinert:2017usi} for $\Lambda =
  500$~MeV. The light- (dark-) shaded bands indicate $95\%$ ($68\%$)
DoB intervals using the Bayesian model $\bar C_{0.5-10}^{650}$. The dotted (dashed) lines show the results based on the
CD Bonn NN  potential \cite{Machleidt:2000ge} (CD Bonn NN potential in combination with
the  Tucson-Melbourne 3NF \cite{Coon:2001pv}). Open circles are
neutron-deuteron from Ref.~\cite{data10_4} and proton-deuteron data from
Ref.~\cite{data10_1,data10_2,data10_3}, corrected for the Coulomb effects, see Ref.~\cite{Epelbaum:2002vt} for
details.  
}
\label{fig:1}       
\end{figure}

\begin{figure}[t]
\includegraphics[width=0.49\textwidth,keepaspectratio,angle=0,clip]{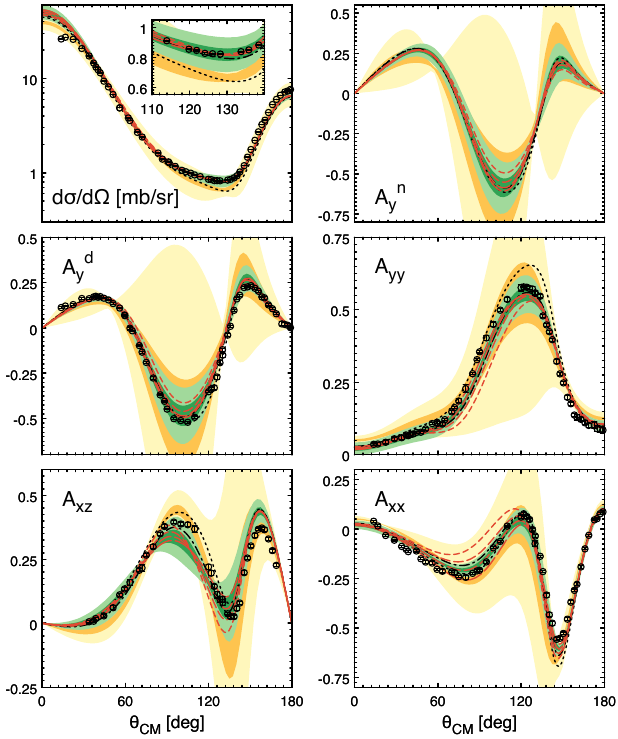}
\caption{Results for the differential cross section, nucleon and
  deuteron analyzing powers $A_y^n$ and $A_y^d$ as well as deuteron
  tensor analyzing powers $A_{yy}$, $A_{xz}$ and $A_{xx}$ in elastic
  nucleon-deuteron scattering  at laboratory energy of
  $E_{\rm lab}^N = 70$~MeV  at NLO
  (yellow bands) and N$^2$LO (green bands) based on the SMS NN
  potentials of Ref.~\cite{Reinert:2017usi} for $\Lambda =
  500$~MeV. Red dashed lines show the
  N$^2$LO results for the cutoff values of $\Lambda = 400$, $450$,
  $500$ and $550$~MeV (the lines with a shorter dash length correspond
  to smaller cutoff values).   
  Open circles are proton-deuteron data from
  Ref.~\cite{Sekiguchi:2002sf}. For remaining notation see
  Fig.~\ref{fig:1}.}
\label{fig:2}       
\end{figure}

\begin{figure}[tb]
\includegraphics[width=0.49\textwidth,keepaspectratio,angle=0,clip]{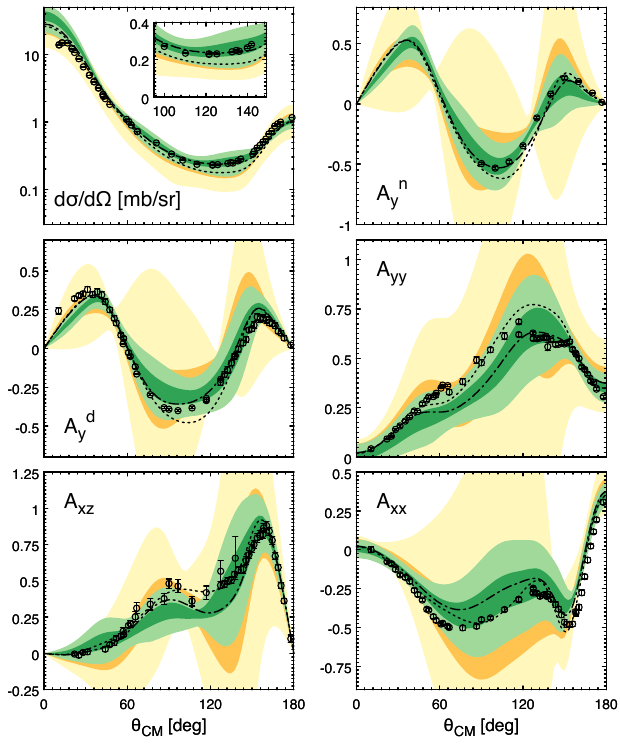}
\caption{Predictions for the differential cross section, nucleon and
  deuteron analyzing powers $A_y^n$ and $A_y^d$ as well as deuteron
  tensor analyzing powers $A_{yy}$, $A_{xz}$ and $A_{xx}$ in elastic
  nucleon-deuteron scattering  at laboratory energy of
  $E_{\rm lab}^N = 135$~MeV at NLO
  (yellow bands) and N$^2$LO (green bands) based on the SMS NN
  potentials of Ref.~\cite{Reinert:2017usi} for $\Lambda =
  500$~MeV. Open circles are proton-deuteron data from
  Ref.~\cite{Sekiguchi:2002sf}. For remaining notation see
  Fig.~\ref{fig:1}.}
\label{fig:3}       
\end{figure}

\begin{figure}[t]
\includegraphics[width=0.49\textwidth,keepaspectratio,angle=0,clip]{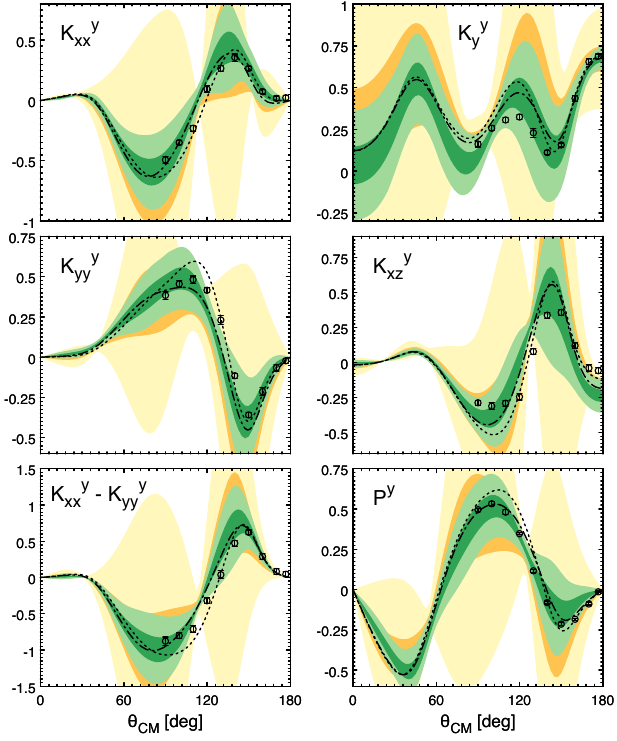}
\caption{Predictions for polarization transfer coefficients
  $K_{xx}^y$, $K_{y}^y$, $K_{yy}^y$, $K_{xz}^y$, $K_{xx}^y-K_{yy}^y$
  and the induced polarization $P^y$  in elastic
  nucleon-deuteron scattering  at laboratory energy of
  $E_{\rm lab}^N = 135$~MeV at NLO
  (yellow bands) and N$^2$LO (green bands) based on the SMS NN
  potentials of Ref.~\cite{Reinert:2017usi} for $\Lambda =
  500$~MeV. Open circles are proton-deuteron data from
  Ref.~\cite{Sekiguchi:2004yb}. For remaining notation see
  Fig.~\ref{fig:1}.}
\label{fig:4}       
\end{figure}

In Figs.~\ref{fig:1}-\ref{fig:4} we show the results for the
differential cross section and selected polarization observables in
elastic nucleon-deuteron scattering
at NLO and N$^2$LO at energies of $E_{\rm lab} = 10$~MeV,  $E_{\rm
  lab} = 70$~MeV and $E_{\rm
  lab} = 135$~MeV for the cutoff $\Lambda = 500$~MeV, along with the
truncation errors corresponding to the DoB intervals of $68\%$ and
$95\%$. 
Notice that the truncation errors are symmetric, and the actual
results of our calculation lie in the middle of the corresponding
error bands.

To estimate the truncation uncertainty we have used 
the Bayesian model $\bar C_{0.5-10}^{650}$ introduced in section
\ref{sec:2}.
For scattering
reactions involving three and more nucleons, we, however, also need to
specify the pertinent momentum $p$ scale that sets the 
expansion parameter $Q$ in Eq.~(\ref{ExpParam}). Consider nucleon-nucleus
scattering and let $E_{\rm lab}^N$ be nucleon beam energy in the laboratory
system. Neglecting the neutron-proton mass difference and the binding
energy of the target nucleus, which is assumed to consist of $A$ nucleons, the CM momentum is related to
$E_{\rm lab}^N$ via 
\beq
\label{pcmelab}
\vec p_{\rm CM}^2 = \frac{E_{\rm lab}^N A^2 m^2 (E_{\rm lab}^N + 2
    m)}{m^2 (A+1)^2 + 2 A m E_{\rm lab}^N} \simeq 
  \frac{2A^2}{(A+1)^2} m E_{\rm lab}^N\,,
\eeq
where $m$ is the nucleon mass and ``$\simeq$'' refers to the nonrelativistic approximation. 
Identifying the scale $p \equiv | \vec p|$ in  Eq.~(\ref{ExpParam})
with $p_{\rm CM} \equiv | \vec p_{\rm CM}  |$
results in $A$-dependent values of the expansion parameter $Q$
corresponding to the same excess energy. For example, the pion
production threshold in the NN (Nd) system with $E_{\rm lab}^N \sim
290$~MeV ($E_{\rm lab}^N \sim 215$~MeV) corresponds to $p_{\rm CM}
\sim 370$~MeV ($p_{\rm CM} \sim 425$~MeV), leading to the expansion
parameter of $Q = 0.57$ ($Q =
0.65$). Alternatively, one can define the momentum scale $p$ in
terms of the Lorentz-invariant excess energy $\sqrt{s} - \sqrt{s_0} =\sqrt{s} - (A + 1) m$ available in the
$A+1$-nucleon system and define the momentum scale $p$ via the relation
\beq
\sqrt{s} - (A + 1) m =: 2 \sqrt{\vec p^2 + m^2} - 2 m\,,
\eeq
that ensures that $p$ coincides with  $p_{\rm CM}$ in the NN
system. Here, $s$ is the usual Mandelstam variable. One can thus
express the scale $p$ in terms of $E_{\rm lab}^N$ via
\beq
\vec p^2 = \frac{s - 2 (A-1) m \sqrt{s} + (A+1) (A-3) m^2}{4}
\eeq
with $s = m^2 (A + 1)^2 + 2 A m E_{\rm lab}^N$. In the nonrelativistic
approximation, this relation simplifies to
\beq
\label{Scalep}
\vec p^2 = \frac{A}{A+1} m E_{\rm lab}^N\,.
\eeq
The nonrelativistic approximation holds at a sub-percent level for the energy range
considered in this study and we use the relation (\ref{Scalep}) to define the expansion
parameter parameter $Q$ in Eq.~(\ref{ExpParam}). 
The breakdown scale $\Lambda_b = 650$~MeV then corresponds to the
excess energy of $\sim 400$~MeV independently of the number of
nucleons $A$ in the target nucleus. Notice that the employed 
model leads to less conservative error estimates for $A>1 $ as compared with
the assignment of $p = p_{\rm CM}$ in Eq.~(\ref{ExpParam}).

We now turn to the results for Nd scattering observables at $E_{\rm
  lab}^N = 10 \ldots 135$~MeV shown in
Figs.~\ref{fig:1}-\ref{fig:4}. Except for the differential cross
section at $E_{\rm lab}^N = 70$~MeV shown in the upper left panel of
Fig.~\ref{fig:2}, the results at N$^2$LO can be regarded as
parameter-free predictions. It is reassuring to see that the calculated observables
are in a reasonably good agreement with the experimental data, which
in most cases lie within the $95\%$ DoB intervals. One should,
however, keep in mind that the estimated truncation errors depend on
the Bayesian model, assumed prior sets and the values of parameters
$M_\pi^{\rm eff}$ and $\Lambda_b$. While model dependence of
uncertainty estimates is expected to decrease at high chiral orders,
it may still be significant at N$^2$LO.

While we only show the uncertainty bands for the cutoff value of
$\Lambda = 500$~MeV, the results for different  values of
$\Lambda$ are similar. To illustrate this point we plot in Fig.~\ref{fig:2}
for the intermediate energy of $E_{\rm lab}^N = 70$~MeV the N$^2$LO results for
all available cutoff values in the range of $\Lambda = 400 \ldots
550$~MeV. Notice that the residual cutoff dependence of the considered
observables at NLO is similar to the one at N$^2$LO and is, in most cases,
comparable with the N$^2$LO $68\%$ DoB intervals. These results
demonstrate that our calculations for different values of $\Lambda$
are consistent with each other within errors. Notice further that the softest
cutoff of $\Lambda = 400$~MeV shows the largest deviation from the
bulk behavior and from the experimental data, which presumably points to
the increasing amount of finite-regulator artifacts.  

We also show in Figs.~\ref{fig:1}-\ref{fig:4} the results based on the
CD Bonn NN potential \cite{Machleidt:2000ge}  with and without the
Tuscon-Melbourne (TM) 3NF \cite{Coon:2001pv}. In particular, for the cases
where the TM 3NF is known to provide sizable corrections such as
e.g.~for the differential cross section around its minimum and for the
deuteron analyzing powers at the intermediate energies of $E_{\rm
  lab}^N = 70$ and $135$~MeV, our N$^2$LO results agree well with the  
CD Bonn NN plus TM 3NF calculations, while the predictions based on the CD
Bonn NN force alone are often outside the $65\%$ and sometimes even
$95\%$ DoB intervals.  This should not come as surprise given the
known weak dependence of the Nd elastic scattering observables on the
off-shell behavior of the NN potentials \cite{Gloeckle:1995jg} and a similar
structure of the TM and the leading chiral 3NF at N$^2$LO
\cite{Epelbaum:2007sq,Meissner:2008zza} largely driven by the intermediate
$\Delta$(1232) excitation.   


Similarly to our
findings in Ref.~\cite{Epelbaum:2018ogq} based on the SCS interactions of
Refs. \cite{Epelbaum:2014efa,Epelbaum:2014sza}, the nucleon and deuteron vector
analyzing 
powers are also properly described (within errors) at the lowest considered energy of 
$E_{\rm lab}^N = 10$~MeV showing no evidence of the so-called
$A_y$-puzzle \cite{Gloeckle:1995jg}  at this chiral order. The much larger
truncation uncertainty for the vector analyzing powers at this low
energy as compared with other observables indicates their strongly
fine-tuned nature, see also Ref.~\cite{Epelbaum:2018ogq} for a related
discussion.
We further emphasize that the
approximate subtraction of the Coulomb effects from the 
proton-deuteron data at this energy may lead to sizable
uncertainties. 

In Refs.~\cite{Binder:2015mbz,Binder:2018pgl}, we have calculated Nd scattering observables based
on the SCS NN chiral potentials of Refs.~\cite{Epelbaum:2014efa,Epelbaum:2014sza} and estimated the
truncation errors using the EKM approach. While these calculations are
incomplete starting from N$^2$LO due to the missing 3NF, they have
demonstrated that the expected accuracy of chiral EFT at high orders such as
N$^4$LO should be substantially smaller than the observed
discrepancies between state-of-the-art calculations and experimental
data.
\begin{figure}[t]
\includegraphics[width=0.49\textwidth,keepaspectratio,angle=0,clip]{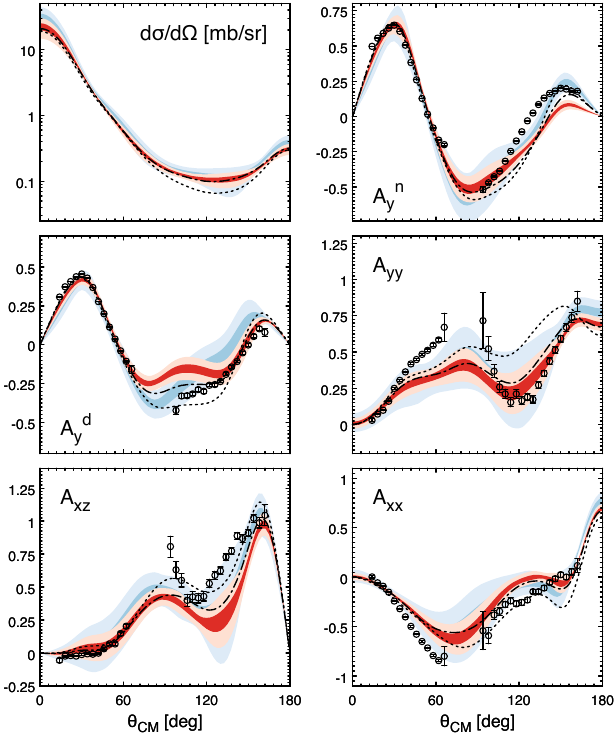}
\caption{Results for the differential cross section, nucleon and
  deuteron analyzing powers $A_y^n$ and $A_y^d$ as well as deuteron
  tensor analyzing powers $A_{yy}$, $A_{xz}$ and $A_{xx}$ in elastic
  nucleon-deuteron scattering  at laboratory energy of
  $E_{\rm lab}^N = 200$~MeV based on the SMS NN
  potentials of Ref.~\cite{Reinert:2017usi} at N$^3$LO (blue shaded
  bands) and N$^4$LO$^+$ (red shaded bands) combined with the 3NF at
  N$^2$LO using $\Lambda = 500$~MeV. Blue (red) shaded bands show the expected truncation
  uncertainty for \emph{complete} N$^3$LO (N$^4$LO) calculations and are 
  obtained by multiplying the N$^2$LO truncation errors for the
  model $\bar C_{0.5-10}^{650}$ with the expansion parameter
  $Q \simeq 0.55$  ($Q^2 \simeq 0.3$).  Open circles are proton-deuteron data from
  Ref.~\cite{vonPrzewoski:2003ig}. 
For remaining notation see
  Fig.~\ref{fig:1}.
}
\label{fig:5}       
\end{figure}
Fig.~\ref{fig:5} shows an update of these finding by using the new
SMS NN potentials of Ref.~\cite{Reinert:2017usi}, including the 3NF at N$^2$LO and
replacing the EKM approach to estimating truncation errors by the
Bayesian model $\bar C_{0.5-10}^{650}$. Specifically, the incomplete
N$^3$LO and N$^4$LO results shown in this figure are based on the
N$^3$LO and N$^4$LO$^+$ NN potentials accompanied with the
N$^2$LO 3NF with the LECs $c_D$ and $c_E$ being readjusted to
the $^3$H binding energy and the differential cross section at
$E_{\rm lab}^N = 70$~MeV in exactly the same way as done at N$^2$LO.  
In the 3NF, we have used  the values of the LECs $c_i$ from Ref.~\cite{Hoferichter:2015tha}
consistent with the NN interactions at the corresponding chiral order,
namely $c_1=
-1.07$~GeV$^{-1}$, $c_3 = -5.32$~GeV$^{-1}$ and $c_4 =
3.56$~GeV$^{-1}$ at N$^3$LO and $c_1=
-1.10$~GeV$^{-1}$, $c_3 = -5.54$~GeV$^{-1}$ and $c_4 =
4.17$~GeV$^{-1}$ at N$^4$LO, subject to the additional shifts of
\beqa
\delta c_1 &=& - \frac{g_A^2 M_\pi}{64 \pi F_\pi^2} \simeq -0.13\mbox{
  GeV}^{-1}\,,\nn
\delta c_3 = - \delta c_4 &=&  \frac{g_A^4 M_\pi}{16 \pi F_\pi^2} \simeq
0.86\mbox{ GeV}^{-1}\,,
\eeqa
generated by the pion loop contributions to the 3NF
at N$^3$LO \cite{Bernard:2007sp}. Since we do not have complete results beyond N$^2$LO, the
error bands in Fig.~\ref{fig:5} are obtained by just
rescaling the corresponding $68\%$ and $95\%$ N$^2$LO DoB intervals.
The incomplete N$^3$LO and N$^4$LO$^+$ results may, of course, still be
regarded as complete N$^2$LO predictions.
At $E_{\rm lab}^N = 200$~MeV, the N$^3$LO uncertainty bands are still
quite sizable indicating that the N$^4$LO contributions to the 3NF could
play a significant role. Thus, fully in line with the
findings of Ref.~\cite{Binder:2015mbz,Binder:2018pgl}, our results suggest that the accurate
description of Nd scattering data below pion production threshold will
likely require the chiral expansion of the 3NF to be pushed to
N$^4$LO. Notice that the accurate and precise description of
neutron-proton and proton-proton data below pion production threshold
also required the chiral expansion of the NN force to be pushed to
N$^4$LO (or even N$^4$LO$^+$) \cite{Reinert:2017usi}.

\section{Subleading short-range 3NF: An exploratory study }
\label{sec:4}

To include the 3NF contributions beyond N$^2$LO one needs to
regularize the corresponding pion loop contributions consistently with
the NN interactions of \cite{Reinert:2017usi} in a chirally symmetric manner, see
Refs.~\cite{CD18-Hermann,EE-NTSE} for discussion.  Such consistently
regularized pion-exchange contributions to the 3NF are not yet
available beyond N$^2$LO. In addition to long- and intermediate-range
interactions generated by pion-exchange diagrams, the chiral 3NF involves
ten purely short-range operators at N$^4$LO, which have been worked out in
Ref.~\cite{Girlanda:2011fh}. In the exploratory study of
Ref.~\cite{Girlanda:2018xrw}, the effects of these subleading
short-range terms are investigated in proton-deuteron scattering
below $E_{\rm lab}^p = 3$~MeV within the hybrid approach based on
phenomenological two- and three-nucleon forces. The authors of
Ref.~\cite{Girlanda:2018xrw} have succeeded to fit the coefficients of
the short-range operators to obtain a good description of experimental
data at $E_{\rm lab}^p = 3$~MeV. However, except for  
fine-tuned observables like the neutron-deuteron doublet scattering
length, which is well known to be correlated with the $^3$H binding
energy, and the analyzing powers $A_y$ and $i T_{11}$, see
Fig.~\ref{fig:1}, the scattering observables at such low energies are
dominated by the NN interaction, and the 3NFs are expected to play a minor 
role \cite{Gloeckle:1995jg}. On the other hand, large discrepancies between theory
and data are observed in Nd elastic scattering at intermediate and
higher energies, where the 3NFs are expected to play a prominent
role \cite{KalantarNayestanaki:2011wz}.   

To explore the role of the subleading short-range 3NF contributions we
choose two out of the ten terms, namely the isoscalar central and spin-orbit interactions
\beqa
 V_{3N} &=& E_1 \,  \vec q_1^2 + i E_7 \, \vec q_1 \times (\vec K_1 - \vec K_2)
 \cdot (\vec\sigma_1 + \vec{\sigma_2}) \nn
 &+&  \mbox{5 permutations}\,, 
 \eeqa
 where $\vec K_i = (\vec p_i ' + \vec p_i )/2$ while $E_1$ and $E_7$
 denote the corresponding LECs. We apply the same nonlocal Gaussian regulator
 as employed in the N$^2$LO short-range part of the 3NF and defined in Eq.~(\ref{RegulSR}), and restrict
 ourselves to the cutoff $\Lambda = 450$~MeV. In this study we do not attempt
 to determine the LECs $E_1$ and $E_7$ from data but explore 
 effects of these 3NF terms for fixed values of these
 LECs. Specifically, in a complete analogy with Eq.~(\ref{DimLessDE}), we express $E_i$ in terms of dimensionless
 coefficients $c_{E_i}$  according to  
\beq
E_i = \frac{c_{E_i}}{F_\pi^4 \Lambda_\chi^3}\,.
\eeq
For the coefficients $c_{E_i}$, we consider in this study the fixed
values of $c_{E_i} = \pm 2$. Based on the variation of the LEC $c_E$,
which, depending on the cutoff $\Lambda = 400 \ldots 550$~MeV, takes the values in the ranges
of $c_E = -0.3 \ldots 1.2$,   $c_E = -0.3 \ldots 0.8$,   $c_E = -0.6
\ldots 0.7$ and $c_E = -0.5 \ldots 0.7$ when using the the NN
potential at orders N$^2$LO, N$^3$LO, N$^4$LO and N$^4$LO$^+$,
respectively, we
expect that the actual values of $c_{E_i}$ should lie well within the
range spanned by $c_{E_i} = \pm 2$. The
adopted naturalness estimates are, however, 
subject to convention-dependent ambiguities\footnote{For
  example, our notation for the spin-orbit term $\propto E_7$ in the 3NF differs from
  the one adopted in Ref.~\cite{Girlanda:2018xrw} by a factor of $1/2$.} and should be taken with care. A more meaningful
and reliable assessment of the natural size of the LECs can be carried
out in  the spectroscopic basis as done in Ref.~\cite{Reinert:2017usi} for the
NN potentials. This, however, would require the inclusion of a complete
set of independent contact operators in the 3NF at N$^4$LO, which goes
beyond the scope of our study. 

\begin{figure}[tb]
\includegraphics[width=0.49\textwidth,keepaspectratio,angle=0,clip]{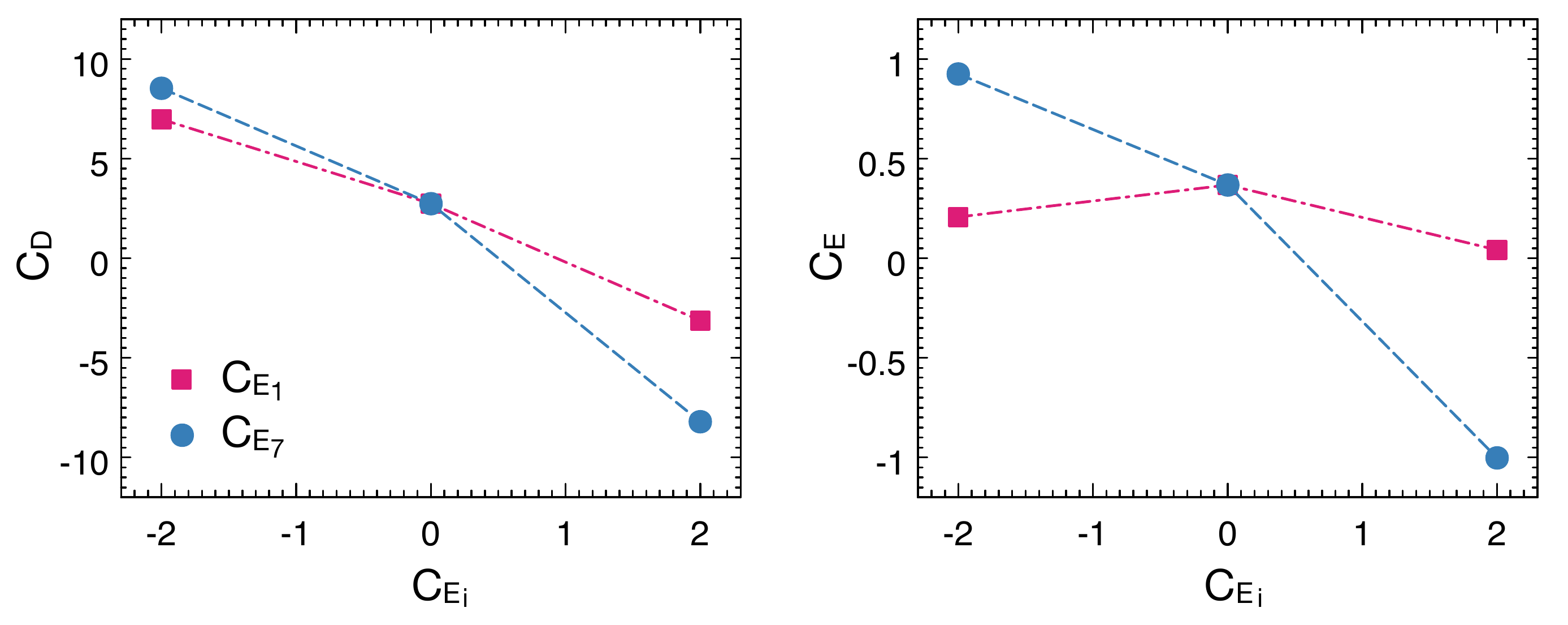}
\caption{LECs $c_D$ and $c_E$ determined from the $^3$H binding
energy and the Nd cross section minimum at $E_{\rm lab}^N = 70$~MeV as
functions of the LECs $c_{E_1}$ (filled purple squares)  and $c_{E_7}$
(filled blue circles).}
\label{fig:X1}       
\end{figure}

For a given observable, the impact of the subleading short-range 3NF
terms can be quantified by comparing the results for  $c_{E_i} = \pm
2$ with those for $c_{E_i} = 0$ after renormalization. Since we are
only able to perform {\it implicit} renormalization by expressing the
bare LECs in terms of low-energy observables, this requires a 
readjustment of the LECs $c_D$ and $c_E$ for every considered set of
$c_{E_1}$, $c_{E_7}$. To allow for a meaningful interpretation of the
obtained results, we follow
in each case exactly the same fitting procedure as explained in
section \ref{sec:3} by adjusting  $c_D$ and $c_E$ to the $^3$H binding
energy and the Nd cross section minimum at $E_{\rm lab}^N = 70$~MeV. 
In Fig.~\ref{fig:X1} we show the resulting dependence of $c_D$ and
$c_E$ on $c_{E_1}$ and $c_{E_7}$. 
In addition to looking at the absolute values of the various
dimensionless LECs, it is also instructive to compare the
corresponding expectation values in the triton state, which are listed
in Table~\ref{ExpVal}. For the considered SMS regulator and $\Lambda =
450$~MeV, the expectation value of the two-pion exchange 3NF
contributions amounts to $\langle V_{2 \pi} \rangle =
-0.19$~MeV.\footnote{We emphasize again that the regularized two-pion
  exchange contributions contain admixtures of the short-range $c_D$-
  and $c_E$-like terms. When adopting the convention of Ref.~\cite{Reinert:2017usi}
  by explicitly subtracting the short-range pieces of the two-pion
  exchange, the expectation value changes to $\langle V_{2 \pi} \rangle =
-0.61$~MeV, showing that the N$^2$LO 3NF is actually dominated by the
long-range pieces.}   
Notice that the spin-orbit 3NF term $\propto c_{E_7}$ does not
contribute to the S-wave partial waves in the triton state
(for the employed angle-independent regulator),
and is found to provide a negligible contribution to the $^3$H binding
energy. The apparent contradiction with the findings of
Ref.~\cite{Girlanda:2018xrw} regarding this term is presumably caused by
a different regulator employed in that paper. 
The expected natural
contribution of the 3NF can be estimated based on naive dimensional
analysis via $| \langle V_{3N} \rangle | \sim Q^3 | \langle V_{2N}
\rangle | \sim 0.3^3 \times 40 \mbox{ MeV} \sim 1$~MeV. For $c_{E_7} =
\pm 2$, the individual terms in the 3NF already start exceeding their
expected natural size, thus indicating that the considered values for
this LEC likely overestimate its natural range. 
This conclusion is supported by Nd scattering results described below. 

We are now in the position to discuss the effects of the
subleading short-range contributions to the 3NF in selected Nd
scattering observables. To that aim, we first perform calculations
based on the NN SMS potential of Ref.~\cite{Reinert:2017usi} at
N$^4$LO$^+$ together with the 3NF at N$^2$LO.

\begin{figure}[tb]
\includegraphics[width=0.49\textwidth,keepaspectratio,angle=0,clip]{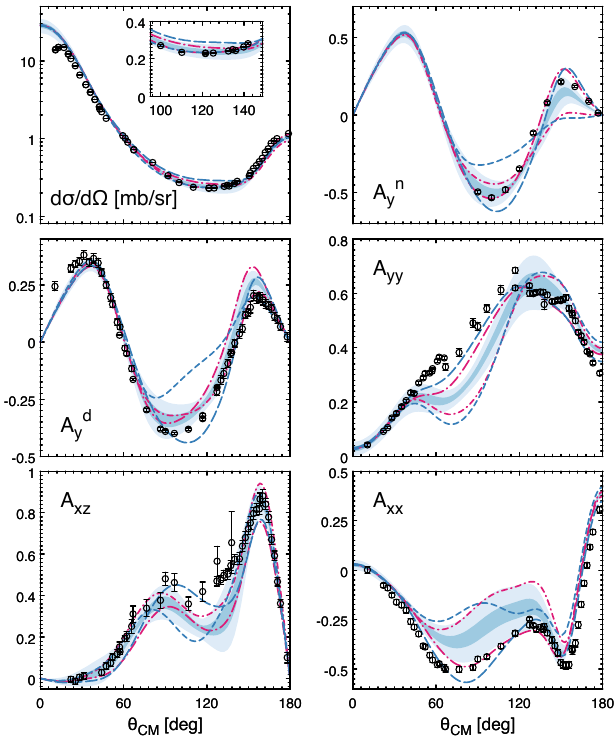}
\caption{Results for the differential cross section, nucleon and
  deuteron analyzing powers $A_y^n$ and $A_y^d$ as well as deuteron
  tensor analyzing powers $A_{yy}$, $A_{xz}$ and $A_{xx}$ in elastic
  nucleon-deuteron scattering  at laboratory energy of
  $E_{\rm lab}^N = 135$~MeV based on the SMS NN
  potentials of Ref.~\cite{Reinert:2017usi} at N$^4$LO$^+$ in
  combination  with the 3NF at
  N$^2$LO using $\Lambda = 450$~MeV. Blue light- (dark-) shaded bands 
 show the expected truncation
  uncertainty for a \emph{complete} N$^3$LO calculation and are
  obtained by multiplying the N$^2$LO truncation error corresponding to
  $95\%$ ($68\%$) DoB intervals for the
  model $\bar C_{0.5-10}^{650}$ with the expansion parameter
  $Q \simeq 0.45$. Short-dashed-dotted and long-dashed-dotted purple lines show the impact
  of the N$^4$LO central short-range 3NF $\propto c_{E_1}$ with $c_{E_1} = -2$ and
  $c_{E_1} = 2$, respectively.  Similarly, short-dashed and
  long-dashed blue lines show the impact
  of the N$^4$LO spin-orbit short-range 3NF $\propto c_{E_7}$ with $c_{E_7} = -2$ and
  $c_{E_7} = 2$, respectively. For remaining notation see
  Fig.~\ref{fig:3}.}
\label{fig:6}       
\end{figure}

\begin{figure}[t]
\includegraphics[width=0.49\textwidth,keepaspectratio,angle=0,clip]{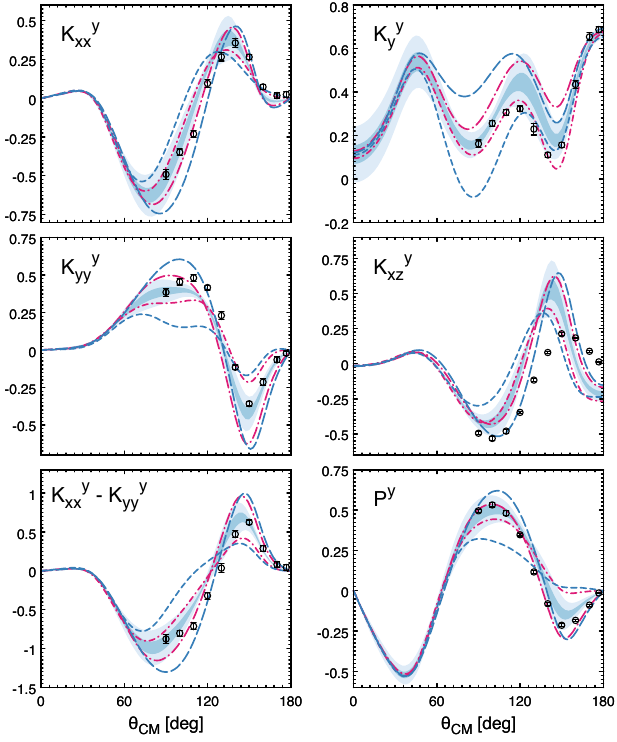}
\caption{Same as Fig.~\ref{fig:X2} but for  polarization transfer coefficients
  $K_{xx}^y$, $K_{y}^y$, $K_{yy}^y$, $K_{xz}^y$, $K_{xx}^y-K_{yy}^y$
  and the induced polarization $P^y$in elastic
  nucleon-deuteron scattering  at laboratory energy of
  $E_{\rm lab}^N = 135$~MeV. For remaining notation see
  Figs.~\ref{fig:3} and \ref{fig:6}.}
\label{fig:7}       
\end{figure}

The resulting predictions for the Nd elastic scattering observables lie in the middle of
the blue bands as shown in Figs.~\ref{fig:6}, \ref{fig:7} for the intermediate
energy of $E_{\rm lab}^N = 135$~MeV. The light- and dark-shaded blue
bands show the $95\%$ and $68\%$ DoB intervals for the truncation
error at N$^3$LO. These error bands do not properly reflect the
uncertainty of our calculation, which is only complete through
N$^2$LO, but show the expected size of N$^4$LO corrections
estimated within the employed Bayesian approach. Next, we repeat the
calculations by switching on the N$^4$LO short-range terms in the
3NF. The resulting predictions for $c_{E_1} = \pm 2$ and $c_{E_7} = 0$
($c_{E_1} = 0$ and $c_{E_7} = \pm 2$) are shown in
Figs.~\ref{fig:6}, \ref{fig:7} by the purple dashed-dotted (blue
dashed) lines. As expected from the estimated truncation uncertainty
at N$^3$LO, the considered N$^4$LO 3NF terms yield sizable
contribution to the Nd scattering observables at 
this rather high energy, especially in the region of the cross section
minimum and at backward angles.
In fact, the magnitude of the $c_{E_1} = \pm 2$
corrections compares well with the width of the
N$^3$LO error bands, especially with the ones corresponding to $95\%$
DoB intervals. This finding supports our expectation that the actual
value of this LEC should be well within the considered range of
$c_{E_1} = \pm 2$. On the other hand, the contributions of the  
 $c_{E_7}$-term lie in most cases outside of the N$^3$LO truncation
 bands, which suggests that the employed values of $c_{E_7}
 = \pm 2$ overestimate the natural size of this LEC. This conclusion
 is in line with the pattern shown in Fig.~\ref{fig:X1}, which
 indicates a significantly larger shifts in the LECs $c_D$,
 $c_E$ induced by changing $\delta c_{E_7} = \pm 2$ as compared with
 the ones induced by $\delta c_{E_1} = \pm 2$.

\begin{table}[tb]
  \caption{Expectation values in the triton state (calculated using the
    N$^4$LO$^+$ NN force alone) of the various short-range terms in
    the 3NF (in MeV) for the cutoff $\Lambda = 450$~MeV.   
\label{ExpVal}}
\begin{tabular*}{0.49\textwidth}{@{\extracolsep{\fill}}crrrr}
   \hline \hline
  &&&&\\[-11pt]
\noalign{\smallskip}
 $(c_{E_1}, \; c_{E_7} )$ &   $\langle V_{D} \rangle $   &  $\langle V_{E} \rangle $   &   $\langle V_{E_1} \rangle $   & 
                                                       $\langle V_{E_7} \rangle $
\smallskip
 \\
  \hline
 &&&&\\[-8pt]  
$(0, \, 0)$  & $0.46$ & $-0.53$&$0$& $0$\\[2pt]
$(2, \, 0)$ & $-0.52$ & $-0.06$ & $0.69$ & $0$\\ [2pt]
$(-2, \, 0)$ & $1.16$
    & $-0.30$ & $-0.69$ & $0$\\ [2pt]
$(0, \, 2)$ & $-1.37$
  & $1.43$ & $0$ & $0$\\ [2pt]
$(0, \, -2)$ & $1.42$
    & $-1.32$ & $0$ & $0$\\ [2pt]  
\hline \hline
\end{tabular*}
\end{table}

Another interesting observation is that both the $c_{E_1}$- and
$c_{E_7}$- contributions  tend to lie well within the truncation bands
at forward angles while outside at backward angles. This could
indicate a shortcoming of the employed Bayesian model,
which relies on Eq.~(\ref{ExpParam}) and does not explicitly account for
higher momentum scales being probed in backward scattering as compared with
forward scattering, see also Ref.~\cite{Melendez:2019izc} for a similar conclusion.


\begin{figure}[tb]
\includegraphics[width=0.49\textwidth,keepaspectratio,angle=0,clip]{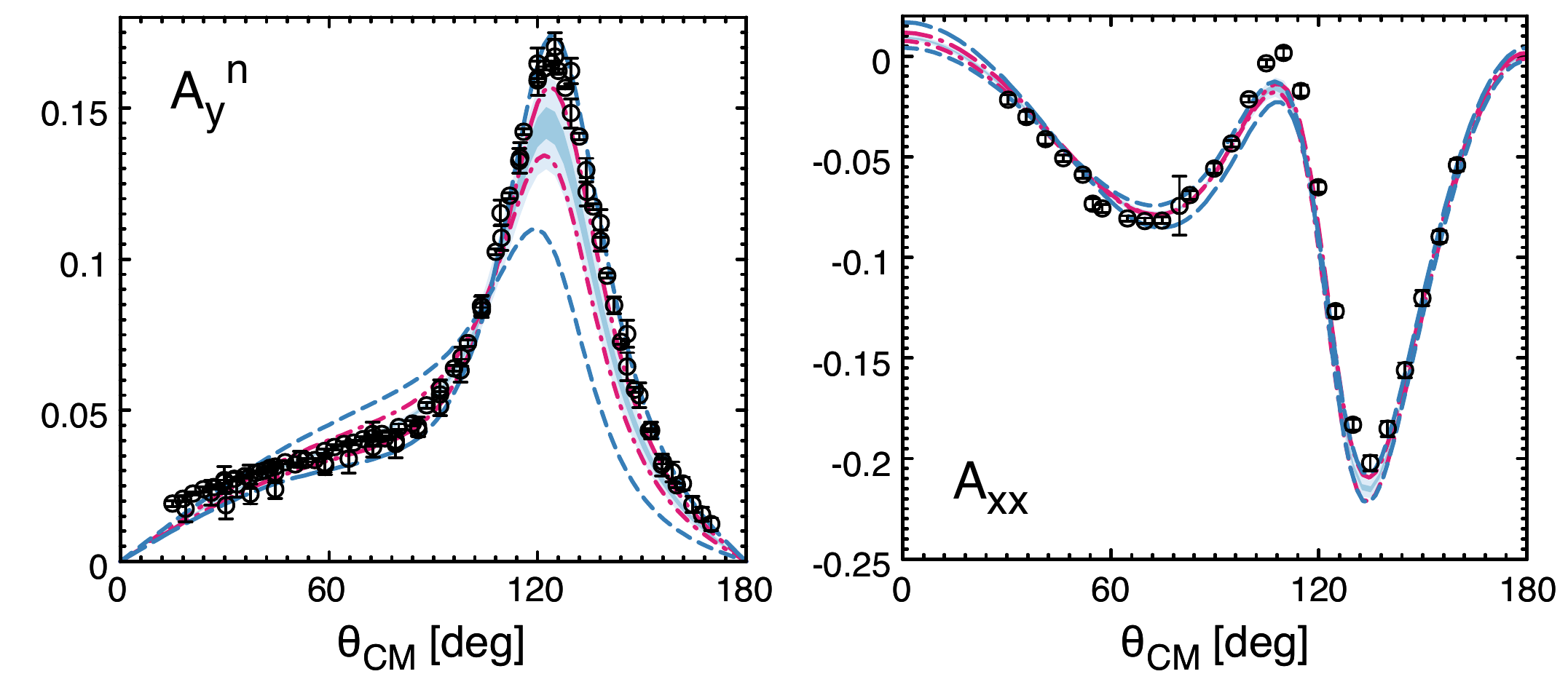}
\caption{Same as Fig.~\ref{fig:6} but for the nucleon vector analyzing power $A_y$ (left
  panel) and deuteron tensor analyzing power $A_{xx}$ (right
  panel) at laboratory energy of
  $E_{\rm lab}^N = 10$~MeV. For remaining notation see
  Figs.~\ref{fig:1} and \ref{fig:6}. }
\label{fig:X2}       
\end{figure}

At the lowest considered energy of $E_{\rm lab}^N = 10$~MeV, the effects of the
considered N$^4$LO 3NF terms turn out to be small, see the right panel of
Fig.~\ref{fig:X2} for a representative example, except for the nucleon and
deuteron vector analyzing powers $A_y^N$ and $A_y^d$. These results
provide yet another confirmation of the fine-tuned nature of these
observables, see the discussion in section \ref{sec:3}, and indicate
that the apparent $A_y$-puzzle could be naturally resolved at the
N$^4$LO level by the corresponding short-range contributions to the
3NF \cite{Girlanda:2016neb},
see also Ref.~\cite{Margaryan:2015rzg} for a related discussion within pionless EFT.
While the spin-orbit 3NF is well known to have a strong impact on the
vector analyzing power \cite{Kievsky:1999nw}, we found, quite surprisingly,
that the isoscalar central short-range 3NF term $\propto c_{E_1}$ 
also significantly affects $A_y$. A comparison of effects due to
the $c_{E_i}$-terms with the estimated truncation error bands at 
$E_{\rm lab}^N = 10$~MeV leads to the same conclusions as reached at
the higher energy of $E_{\rm lab}^N = 135$~MeV. 

Last but not least, we have also looked at the differential cross
section in the so-called symmetric space star configuration of the Nd
breakup reaction at $E_{\rm lab}^N = 13$~MeV, which is known to
represent another low-energy puzzle in the three-nucleon
continuum. Contrary to  $A_y$, this observable is dominated by the
S-wave components of the NN force and turns out to be highly 
insensitive to the 3NFs considered so far, see Ref.~\cite{Witala:2019ffj} for recent
results based on the SCS nuclear potentials of Refs.~\cite{Epelbaum:2014efa,Epelbaum:2014sza,Epelbaum:2018ogq}. We found
that the inclusion of the $c_{E_i}$-terms has a negligible effect on
the cross section in this breakup configuration, and the obtained results
essentially coinside with the one presented in Ref.~\cite{Witala:2019ffj}. The
observed discrepancy between the theoretical calculations and the
neutron- and proton-deuteron data thus indeed appears to be
puzzling. It will be interesting to see if this puzzle can be resolved
by the inclusion of the N$^3$LO and remaining N$^4$LO contributions to
the 3NF.

\section{Summary and conclusions}
\label{sec:5}

In this paper we analyzed selected Nd scattering observables at
N$^2$LO in chiral EFT based on the SMS interactions of
Ref.~\cite{Reinert:2017usi}.
The main results of our study can be summarized as follows. 
\begin{itemize}
\item
Following the approach of Ref.~\cite{Melendez:2017phj}, we have
explored  several pointwise Bayesian models for quantifying  truncation
uncertainties in chiral EFT and tuned them by calculating angular
distributions of neutron-proton scattering.  
\item
Using the SMS NN forces of Ref.~\cite{Reinert:2017usi} accompanied
with the N$^2$LO
3NF regularized in the same way, we have determined
the LECs $c_D$ and $c_E$ entering the 3NF from the $^3$H binding
energy and the Nd cross section data of Ref.~\cite{Sekiguchi:2002sf} at $E_{\rm lab}^N
= 70$~MeV. The resulting N$^2$LO results for elastic Nd scattering
observables agree within errors with our earlier N$^2$LO calculations
based on the SCS interactions \cite{Epelbaum:2018ogq} and with experimental data.
\item
The truncation errors for various Nd scattering observables estimated in
\cite{Binder:2015mbz,Binder:2018pgl,Epelbaum:2018ogq} using the
approach of Ref.~\cite{Epelbaum:2014efa} are found to be consistent 
with $68\%$ DoB intervals for the employed Bayesian model $\bar
C_{0.5-10}^{650}$. In particular, we confirm our earlier findings in
Refs.~\cite{Binder:2015mbz,Binder:2018pgl}, obtained using the SCS NN
forces of Refs.~\cite{Epelbaum:2014efa,Epelbaum:2014sza}, 
that Nd scattering at intermediate and higher energies
provides a ``golden window'' to study higher-order contributions to
the chiral 3NF.
\item
Based on the estimated truncation errors, we argue that a
high-precision description of neutron-deuteron scattering below pion
production threshold will likely require pushing the chiral expansion
of the 3NF at least to N$^4$LO. This conclusion is in line with the
convergence pattern of the chiral expansion in the NN sector \cite{Reinert:2017usi}
and is further substantiated by an exploratory study of the
short-range 3NF contributions $\propto E_{1,7}$, which, as expected
from our Bayesian analysis, are found to have significant effect on Nd
polarization observables at intermediate energies.
Our results show that the contributions to
the 3NF beyond N$^2$LO are indeed potentially capable of resolving the discrepancies
between theory and experiment observed in Nd scattering at
intermediate and higher energies \cite{KalantarNayestanaki:2011wz}. 
\end{itemize}
As a next step, this study should be extended to N$^3$LO, which will
require the inclusion of  {\it consistently
  regularized} 3NF contributions. Work along these lines is in progress
by the LENPIC Collaboration.


\section*{Acknowledgments}
This study has been performed within Low Energy Nuclear Physics
International Collaboration (LENPIC) project and 
was  supported by BMBF
(contract No.~05P18PCFP1),
by DFG and
NSFC (TRR 110), by DFG (project No. 279384907 - CRC 1245),
by the Polish National Science Centre 
under Grants No.~2016/22/M/ST2/00173 and 2016/ 21/D/ST2/01120, 
by the Chinese Academy of Sciences (CAS) Presidents
International Fellowship Initiative (PIFI) (Grant No.~2018DM0034)
and by VolkswagenStiftung (Grant No. 93562). 
Numerical
calculations were performed on the supercomputer cluster of the JSC,
J\"ulich.  

\appendix
\section{Analytic expressions for the posterior pdf}
\label{sec:app1}

For the Gaussian prior pdf in Eq.~(\ref{posteriorGeneral}), the posterior pdf takes
the form 
\beqa
  \label{TruncationErrorFinal}
&& {\rm pr}_h^C ( \Delta | \{ c_{i \le k}\} ) = \frac{1}{\sqrt{\pi \bar q^2
    \fet c_k^2}} \left( \frac{\fet c_k^2}{\fet c_k^2 +
    \Delta^2 / \bar q^2} \right)^{k/2} \\
&& \times \frac{{\rm \Gamma} \left[\frac{k}{2},
  \, \frac{1}{2 \bar c_>^2} \left(  \fet c_k ^2 +
    \frac{\Delta^2}{\bar q^2} \right) \right] - {\rm \Gamma} \left[\frac{k}{2},
  \, \frac{1}{2 \bar c_<^2} \left(  \fet c_k^2 +
    \frac{\Delta^2}{\bar q^2} \right) \right] 
}{{\rm \Gamma} \left[\frac{k-1}{2},
  \, \frac{\fet c_k^2}{2 \bar c_>^2} \right] -  {\rm \Gamma} \left[\frac{k-1}{2},
  \, \frac{\fet c_k^2}{2 \bar c_<^2} \right] }\,,
\nonumber
\eeqa
  where $\bar q^2 \equiv \sum_{i = k+1}^{k+h} Q^{2i}$, $\fet c_k^2
  \equiv \sum_{i\in A} c_i^2$ and the incomplete gamma function is defined as
  \begin{equation}
{\rm \Gamma} (s, x ) = \int_x^\infty dt \, t^{s-1} \, e^{-t}\,.
\end{equation}    
For the noninformative prior $C_\epsilon$ with $\bar c_< = \epsilon$
and $\bar c_> = 1/\epsilon$, the expression for the
posterior, after taking the limit $\epsilon \to 0$ for $\fet c_k^2 \ne
0$ simplifies to
\beq
 {\rm pr}_h^{C_\epsilon} ( \Delta | \{ c_{i \le k}\} ) = \frac{1}{\sqrt{\pi \bar q^2
    \fet c_k^2}} \frac{{\rm \Gamma} \left(\frac{k}{2}\right)}{{\rm \Gamma} \left(\frac{k-1}{2}\right)} \left( \frac{\fet c_k^2}{\fet c_k^2 +
    \Delta^2 / \bar q^2} \right)^{k/2} \,.
\label{priorepsilon}
\eeq

\section{Partial wave decomposition of the N$4$LO 3NF contact terms}
\label{sec:app2}

    For the $E_1$-term
    \beq
    V_{3N} = E_1 \sum_{i \ne j \ne k} \vec q_i^2\,,
    \eeq
    we choose the Faddeev component $V_{3N}^{(1)} = 2\, E_1\, \vec
    q_1^2$ invariant with respect to the interchange of nucleons $2$
    and $3$ and obtain
    \begin{align*}
  &   \langle p' k' \alpha' \vert V_{3N}^{(1)} \vert p k \alpha \rangle = 32 \pi^2\, E_1\, \delta_{s's}\, 
    \delta_{l'0}\, 
    \delta_{l0}\, \delta_{sj'}\, \delta_{sj}\, \delta_{T'T}\, \delta_{M_T'M_T}\, \\
    & \times  \delta_{t't} \left[ (k^2+k'^2)\, \delta_{\lambda'0}\, \delta_{\lambda 0}\, \delta_{I'\frac{1}{2}}\, \delta_{I\frac{1}{2}} - 
    \frac{2}{3}\,k\,k'\, \delta_{\lambda'1}\, \delta_{\lambda 1}\, \delta_{I'I}\right]\,,
    \end{align*}
    with $\vec p$, $\vec p '$, $\vec k$ and $\vec k '$ denoting the
    corresponding Jacobi momenta.  
    
    For the $E_7$-term
    \beq
    V_{3N} = i E_7 \sum_{i \ne j \ne k} \vec q_i \times (\vec K_i - \vec K_j) \cdot (\vec\sigma_i + \vec{\sigma_j})\,,
    \eeq
    we choose the Faddeev component
\beqa
    V_{3N}^{(1)} &=& i \, E_7 \big[\vec q_1 \times (\vec K_1 - \vec
    K_2) \cdot (\vec\sigma_1 + \vec\sigma_2) \nn
    &+& \vec q_1 \times (\vec
      K_1 - \vec K_3) \cdot (\vec\sigma_1 + \vec\sigma_3)\big]
    \eeqa
    and obtain
    \begin{align*}
 &   \langle p' k' \alpha' \vert V_{3N}^{(1)} \vert p k \alpha \rangle
   = 8\pi^2 E_7 \, \delta_{T'T} \, \delta_{M_T'M_T} \, \delta_{t't} \\
      & \times \bigg[- \sqrt{2} \left(1 - (-1)^{s'+s}\right) (-1)^{J+\tfrac{1}{2}}
    \\
    &\times \bigg(p \, k \, \delta_{l'0} \, \delta_{\lambda'0} \, \delta_{l1} \, \delta_{\lambda 1} \, \delta_{s'j'} \, \delta_{I'\tfrac{1}{2}} \, \sqrt{\hat{j}\hat{I}} \,
    \begin{Bmatrix}
    1 & s & s' \\[2pt]
    j & 1 & 1\\
    \end{Bmatrix}
    \begin{Bmatrix}
    s'  & j & 1 \\[2pt]
    I & \tfrac{1}{2} & J\\
    \end{Bmatrix}
    \\
    &-p' \, k \, \delta_{l'1} \, \delta_{\lambda'0} \, \delta_{l0} \, \delta_{\lambda 1} \, \delta_{sj} \, \delta_{I'\tfrac{1}{2}} \, \sqrt{\hat{j}'\hat{I}} \,
    \begin{Bmatrix}
    1 & s' & s \\[2pt]
    j' & 1 & 1\\
    \end{Bmatrix}
    \begin{Bmatrix}
    j' & s & 1 \\[2pt]
    I & \tfrac{1}{2} & J\\
    \end{Bmatrix}
    \\
    &-p \, k' \, \delta_{l'0} \, \delta_{\lambda'1} \, \delta_{l1} \, \delta_{\lambda 0} \, \delta_{s'j'} \, \delta_{I\tfrac{1}{2}} \, \sqrt{\hat{j}\hat{I}'} \,
    \begin{Bmatrix}
    1 & s & s' \\[2pt]
    j & 1 & 1\\
    \end{Bmatrix}
    \begin{Bmatrix}
    j & s' & 1 \\[2pt]
    I' & \tfrac{1}{2} & J\\
    \end{Bmatrix}
    \\
    &+p' \, k' \, \delta_{l'1} \, \delta_{\lambda'1} \, \delta_{l0} \, \delta_{\lambda 0} \, \delta_{sj} \, \delta_{I\tfrac{1}{2}} \, \sqrt{\hat{j}'\hat{I}'} \,
    \begin{Bmatrix}
    1 & s' & s \\[2pt]
    j' & 1 & 1\\
    \end{Bmatrix}
    \begin{Bmatrix}
    s  & j' & 1 \\[2pt]
    I' & \tfrac{1}{2} & J\\
    \end{Bmatrix}\bigg)
    \\
    &+12\,  k' \, k \, \delta_{l'0} \, \delta_{\lambda'1} \, \delta_{l0} \, \delta_{\lambda 1} \, \delta_{s's} \, \delta_{s'j'} \, \delta_{sj} \bigg( \delta_{I'I}(-1)^{I+\tfrac{1}{2}} 
    \begin{Bmatrix}
    \tfrac{1}{2} & 1 & I\\[2pt]
    1 & \tfrac{1}{2} & 1
    \end{Bmatrix}
    \\
    &+ \delta_{s1}\sqrt{\hat{I}'\hat{I}} \, (-1)^{I'+I+J+\tfrac{1}{2}}
    \begin{Bmatrix}
    1 & I & I'\\[2pt]
    \tfrac{1}{2} & 1 & 1
    \end{Bmatrix}
    \begin{Bmatrix}
    1 & I & I'\\[2pt]
    J & 1 & 1
    \end{Bmatrix}
    \bigg)\bigg]\,,
    \end{align*}
with $\hat X \equiv 2 X + 1$. For more details on notation see
Ref.~\cite{Gloeckle:1995jg}.

\end{document}